\documentclass[twocolumn]{aastex62}

\usepackage{amsmath, amsthm, amssymb}
\usepackage{pbox}
\usepackage{booktabs,tabularx}
\usepackage{multirow}
\usepackage{graphicx}
\usepackage[flushleft]{threeparttable}
\usepackage{enumitem}

\usepackage[utf8]{inputenc}
\usepackage[T1]{fontenc}

\graphicspath{{./}{figures/}}

\submitjournal{ApJ}
\accepted{}

\shorttitle{Clouds in \textit{JWST} transmission spectra}
\shortauthors{Mai \& Line}

\begin{document}

\title{Exploring Exoplanet Cloud Assumptions in \textit{JWST} Transmission Spectra}

\correspondingauthor{Chuhong Mai}
\email{chuhong.mai@asu.edu}

\author[0000-0002-9243-5065]{Chuhong Mai}
\affil{School of Earth and Space Exploration, Arizona State University, Tempe, AZ 85287}
\author{Michael R. Line}
\affiliation{School of Earth and Space Exploration, Arizona State University, Tempe, AZ 85287}




\begin{abstract}
Clouds are ubiquitous in extrasolar planet atmospheres and are critical to our understanding of planetary climate and chemistry. They also represent one of the greater challenges to overcome when trying to interpret transit transmission spectra of exoplanet atmospheres as their presence 
can inhibit precise constraints on atmospheric composition and thermal properties. 
In this work we take a phenomenological approach towards understanding 1) our ability to constrain bulk cloud properties, and 2) the impact of clouds on constraining various atmospheric properties as obtained through transmission spectroscopy with the \textit{James Webb Space Telescope (JWST)}. 
We do this by exploring retrievals of atmospheric and cloud properties for a generic ``hot-Jupiter'' as a function of signal-to-noise ratio (SNR), \textit{JWST} observing modes and four different cloud parameterizations.
We find that most key atmospheric and cloud inferences can be well constrained in the wavelength range ($\lambda = $ 0.6 - 11 $\mu$m), with NIRCam ($\lambda =$ 2.5 - 5 $\mu$m) being critical in inferring atmospheric properties and NIRISS + MIRI ($\lambda =$ 0.6 - 2.5, 5 - 11 $\mu$m) being necessary for good constraints on cloud parameters. 
However, constraining the cloud abundance and therefore the total cloud mass requires an observable cloud base in the transit geometry.
While higher SNR observations can place tighter constraints on major parameters such as temperature, metallicity and cloud sedimentation, they are unable to eliminate strong degeneracies among cloud parameters.
Our investigation of a generic ``warm-Neptune'' with photochemical haze parameterization also shows promising results in constraining atmospheric and haze properties in the cooler temperature regime.

\end{abstract}



\section{Introduction} \label{sec:intro}

Transmission spectroscopy has been widely applied to characterize the terminator properties of exoplanet atmospheres, including atmospheric composition, thermal structure, dynamics, and climate etc (e.g., \citealt{Seager2000, Swain2008, Huitson2012}). A particularly intriguing find among these inferences is the ubiquitous presence of clouds and hazes in these alien atmospheres, which has been believed to be the most likely cause of weakened spectral features in transmission spectra (e.g., \citealt{Crossfield2013, Kreidberg2014, Knutson2014a, Knutson2014b, Iyer2016, Sing2016}). \cite{Sing2016} showed that there exists a continuum in the broad-band (0.3 - 5 $\mu$m) transmission spectra feature morphology of ten hot-Jupiter's, ranging from large features assumed to arise from predominately clear atmospheres to muted features with gentle power-law like slopes in the optical range indicative of the presence of obscuring clouds and/or hazes. Observations of smaller, cooler planets within the ``Neptune-to-Super-Earth'' regime have revealed predominately flat or nearly flat spectra over the optical to near-infrared range\citep{Kreidberg2014, Knutson2014a, Knutson2014b, Fraine2014}.

It is vital to our understanding of exoplanetary atmospheres to determine, at a minimum, the influence that clouds and/or hazes will have on our ability to infer basic atmospheric properties.
Given the large parameter space of clouds/hazes and possible degeneracies among parameters, the modeling of clouds/hazes has been difficult. 
To date, many cloud models for substellar atmospheres have been formulated and developed to help us better understand the microphysical formation of clouds and hazes, and to assist the interpretation of spectral data.

These models can be categorized into two main groups. The first is sophisticated cloud/haze models that describe full microphysical schemes for cloud/haze particulate formation, growth, sedimentation and mixing self-consistently (e.g. \citealt{Helling2008a, Helling2008b, Allard2012, Lavvas2017, Kawashima2018, Ohno2018, Gao2018a, Gao2018b, Powell2018}).
3-D simulations have been performed utilizing these microphysical cloud models for substellar atmospheres (e.g. \citealt{Parmentier2013, Lee2015, Lee2016, Helling2016, Lines2018}).

The second group includes parametric cloud/haze models, typically 1D, to describe cloud/haze particle distributions without the inclusion of complex microphysics (e.g. \citealt{Ackerman2001, Tsuji2002, Burrows2006, Marley2010, Madhusudhan2011, Morley2012, Morley2014, Charnay2018, Ormel2019}).
The latter have been more commonly used due to their simplicity, and fewer tunable parameters.


Numerous past investigations have studied the impact of clouds/hazes on transmission spectra (e.g., \citealt{desEtangs2008, Morley2013, Line2016, Wakeford2017, Pinhas2017}). 
Obtaining a detailed understanding of cloud properties has proven difficult due to the inherent degeneracies that present themselves when data are limited in a spectral range \citep{Tsiaras2018}.  
\textit{The James Webb Space Telescope (JWST)}, with its broader wavelength coverage, higher photometric precision and resolution, is anticipated to provide greater sensitivity to cloud properties like particle sizes, vertical distributions, and cloud composition (e.g., \citealt{Pinhas2017, Lines2018}). 

Atmospheric retrieval has proven to be a useful tool for quantitatively assessing the degeneracies in transit spectra and identifying nominal observation strategies to break them (e.g., \citealt{Benneke2013, Line2013, Greene2016, Feng2018}).  For instance, \citet{Greene2016} explored the potential for different \textit{JWST} instruments (NIRISS, NIRCam, and MIRI) to constrain the molecular abundances on a variety of planet types, from hot-Jupiter's to cool sub-Neptune's.  \cite{Rocchetto2016} explored the impact of assuming isothermal atmospheres on compositional inferences.  \cite{Feng2016} and \cite{Blecic2017} determined how 3D effects would bias 1D thermal emission retrievals.  

Here, we extend the work of \citet{Greene2016} to explore the constraints obtainable on cloud properties and how differing assumptions in cloud models under anticipated \textit{JWST} observational scenarios will influence our atmospheric inferences.  We aim to address the following questions:  
\begin{itemize}
\item How well can we constrain the properties of clouds and the atmospheric composition under different observation conditions of \textit{JWST} transmission spectra (e.g. different noise levels, different instrument modes)?
\item How do different cloud parameterizations affect these constraints? Do they provide different interpretations of clouds for the same spectral data?

\item Do cloud assumptions bias atmospheric inferences? If yes, how?
\end{itemize}


Our paper is organized as follows: \S~ \ref{sec:models} describes the model set up, including the cloud/haze models adopted (\S~ \ref{subsec:cloudmodels}), simulations of \textit{JWST} transmission spectra with the forward model (\S~ \ref{subsec:forward}), 
the retrieval forward model parameterization and experimental setup (\S~ \ref{subsec:retrieval}). 
\S~ \ref{sec:results} presents the results from these experiments with different observation conditions (\S~ \ref{subsec:res_obs}) and cloud parameterizations (\S~ \ref{subsec:res_cloud}). 
Outstanding degeneracies found among cloud parameters (\S~ \ref{subsec:degen}) and the constraint on cloud mass (\S~ \ref{subsec:cloudmass}) are also discussed.
We discuss our findings in \S~ \ref{sec:discus} and summarize the results in \S~ \ref{sec:sum}.

\section{Model Set-Up} \label{sec:models}
The primary objectives of this work are to determine the degree to which we can constrain cloud/haze properties under different model parameterizations with different \textit{JWST} observing mode setups, as well as how different cloud assumptions could bias constraints on other fundamental atmospheric properties like metallicity ([Fe/H]) or carbon-to-oxygen ratio (C/O).  
The main approach is to generate simulated \textit{JWST} observations and retrieve upon them under different cloud model assumptions.

Cloud composition and other properties are anticipated to wildly vary across the broad range of planetary conditions (temperatures, gravities, irradiation levels, etc. \citealt{Morley2013, Pinhas2017}). To simplify this challenge we studied two planet archetypes: a generic hot-Jupiter, based upon the WASP 62 system, and a warm-Neptune modeled after the GJ 1214 system.  In the hot-Jupiter scenario we set up a comprehensive simulation grid in signal-to-noise ratio (SNR, represented by different scalings of error bars in this work), and different observation modes/wavelength coverages following those choices in \citet{Greene2016}.
These simulated synthetic spectra are then retrieved upon in order to determine the biases and degeneracies within our various choices of cloud models (\S~\ref{subsec:cloudmodels}).


\subsection{Cloud Models} \label{subsec:cloudmodels}
To test how the hierarchy of cloud assumptions affects our ability to constrain atmospheric and cloud quantities in \textit{JWST} data, we investigate four cloud models for the hot-Jupiter in this study (summarized in Table \ref{tab:models}).  For mie scattering models we assume a single condensate composition made of enstatite (MgSiO$_3$) with the total extinction coefficient computed using the {\tt PyMieCoated} routine\footnote{https://github.com/jleinonen/pymiecoated/} given the particle size and the indices of refraction from Table 1 of \citet{Wakeford2015}. We assume pure extinction only (e.g., no Monte Carlo scattering). For parameterizations that include cloud particle sizes and/or distributions, we include particles with sizes ranging from 10$^{-3}$ $\mu$m to 10$^3$ $\mu$m, equally divided into 61 size bins ($\sim$ 10 bins for each radius decade). 


\textbf{The Power Law Haze + Gray cloud model}, is the most overly-simplistic but most widely-used model within retrieval parameterizations.  It features an ad-hoc parameterization of cloud top pressure ($P_c$) and haze cross-section ($\sigma$) \citep{desEtangs2008, Greene2016}:
\begin{equation} \label{eq1}
\sigma = \sigma_0\left(\frac{\lambda}{\lambda_0}\right)^{-\beta}
\end{equation}
with $\sigma_0$ being the amplitude of the haze cross-section relative to H$_2$ Rayleigh scattering at 0.4 $\mu$m, and $\beta$ describes the scattering slope (with $\beta$=4 as Rayleigh scattering).
As in numerous previous works, the cloud top pressure, $P_c$, simply flattens transmission spectra by setting the limb transmittances deeper than the cloud-top-level to zero, while $\sigma_0$ and $\beta$ shape the power law slope at shorter wavelengths.  These parameters contain virtually no physics and are simply used to encapsulate/marginalize over the ``continuum shape'' of seemingly cloudy/hazy spectra. Surprisingly, such a simplistic parameterization adequately explains the broad band nature of most observed transmission spectra (e.g., \citealt{Sing2016}).   

\textbf{The One-size cloud model} assumes a vertically distributed cloud composed of single-size mie scattering droplets/ particles (no size distributions). The model is parameterized with a droplet mixing ratio at the cloud base, a single particle radius, a cloud base pressure, and a vertical profile index. Note that all mixing ratios mentioned in this paper are volume mixing ratios unless specified. The vertical distribution of the cloud droplet mixing ratio is given by:
\begin{equation} \label{eq2}
  F_{drop}=F_{drop,0}\left(\frac{p}{p_0}\right)^{\alpha}
\end{equation}
where $F_{drop}$ is the local mixing ratio of droplets at pressure $p$ and $F_{drop,0}$ is the mixing ratio at the cloud base pressure $p_0$. 
The vertical index, $\alpha$, describes how the droplet mixing ratio falls off with pressure. 
$\alpha=0$ means the cloud particles are so well mixed that the droplet mixing ratio stays constant throughout the cloud layer. 
Such a condition typically requires an extremely turbulent atmosphere or the cloud particles to be small and light enough to be easily transported.
$\alpha>0$ causes the condensate mixing ratio to fall off with altitude (e.g., due to sedimentation). 
In typical atmospheric environments, the bigger the particle size, the larger $\alpha$ becomes. 

\textbf{The Integrated cloud model} is slightly more physically motivated parameterization when compared to the One-size model, in that it takes the particle size distribution into consideration. 
Instead of assigning an independent parameter to describe the vertical profile, this model assumes an approximate relationship between particle radius ($r_{drop}$) and $\alpha$ based on results in \cite{Parmentier2013} (Fig. 4 and Fig. 10 in their work):
\begin{equation} \label{eq3}
  \alpha = ar_{drop}^b+c
\end{equation}
where $a=0.005, b=4.90,c=-0.13$ are the fitting coefficients.
At the cloud base, the abundances of cloud particles with different sizes are determined by a log-normal size distribution with a fixed geometric standard deviation (e.g. $\sigma$ = 2). 
Each particle size follows a specific vertical profile described by its $\alpha$ calculated in Eqn. \ref{eq3}.
Their mixing ratios are then calculated through Eqn. \ref{eq2} for each pressure level.
As such, small particles become more dominant at higher altitudes. The size distribution of particles at these altitudes does not maintain a log-normal shape, but rather, has a gentle cutoff in the wing of large particles. 
Using only three parameters - the droplet mixing ratio at the cloud base, the mean particle radius at cloud base and the cloud base pressure, the Integrated cloud model intends to capture a more realistic picture of clouds.

\textbf{The Ackerman \& Marley cloud model}, hereafter \textbf{the A \& M model}, was first proposed by \cite{Ackerman2001}, with following developments in \cite{Marley2010}, \cite{Morley2012, Morley2014} and \cite{Charnay2018}. 
The model calculates cloud profiles by balancing the upward vertical mixing and downward sedimentation of cloud particles:
\begin{equation} \label{eq4}
  -K_{zz}\frac{\partial (q_c+q_v)}{\partial z}-v_{sed}\cdot q_c = 0
\end{equation}
where $q_c$ and $q_v$ are the mass mixing ratios of condensates and vapor. $K_{zz}$ is the eddy diffusion coefficient used to describe the level of vertical mixing in the atmosphere by turbulence or convection. $K_{zz}$ is assumed as constant throughout the atmosphere in our adopted model.
$v_{sed}$ is the particle sedimentation velocity. 
A more convenient way to parameterize sedimentation is to use the ratio of particle sedimentation velocity to the characteristic vertical mixing velocity $f_{sed} = v_{sed}/w^*$, where the vertical mixing velocity $w^*=K_{zz}/H$ \citep{Charnay2018}. 
The A \& M model assumes a constant $f_{sed}$ value throughout the model to determine the vertical profiles of condensates with different particle sizes.
$f_{sed}$ plays a similar role as $\alpha$ (Eqn. \ref{eq2}) in sculpting the vertical cloud profile -- a larger $f_{sed}$ represents stronger sedimentation and thus more compact clouds. 
Both $f_{sed}$ and $K_{zz}$ set the nominal droplet size in each cloud layer (modified from Equation 10 in \cite{Charnay2018}):

\begin{equation} \label{eq5}
r_{sed}=\frac{2}{3}\times \lambda \left( \sqrt[]{1+10.125\frac{\eta K_{zz}f_{sed}}{gH(\rho_p-\rho_a)\lambda^2}}-1 \right)
\end{equation}
where $r_{sed}$ is the sedimentation radius of cloud particles, the equivalent radius of mass-weighted sedimentation flux; $\lambda$ is the mean free path; $\eta$ is the gas viscosity; $g$ is gravity; $H$ is the scale height of the atmosphere and $\rho_p$ and $\rho_a$ are the densities of the particles and the atmosphere.
The model then distributes particles to follow the log-normal size distribution centering their mean sizes (derived from $r_{sed}$ using equation 13 in \citet{Ackerman2001}) in each layer. 
As in \cite{Charnay2018} we assumed that contribution due to condensation between two cloud levels is negligible (due to rapid rainout) compared to advection, allowing us to to specify the condensate (pre-droplet) mixing ratio at the cloud base ($f_{cond}$), and the condensate mixing ratio profile. Combining hydrostatic equilibrium with term ``A'' (equation 3) in \cite{Charnay2018}, along with the definition of sedimentation velocity and $f_{sed}$, the cloud vertical profile can be prescribed with
\begin{equation}\label{eq:am_profile}
    f(P)=f_{cond} (P/P_{base})^{f_{sed}}
\end{equation}
for $P<P_{base}$.

This cloud model self-consistently determines the cloud-droplet vertical distribution given the condensate mixing ratio at cloud base, the cloud base pressure, $f_{sed}$, and $K_{zz}$.  The A \& M cloud model is the fiducial/baseline cloud model used in the hot-Jupiter simulations. 

\begin{figure*}[ht!]
\centering
\hspace*{-1.2cm}
\includegraphics[scale=0.3]{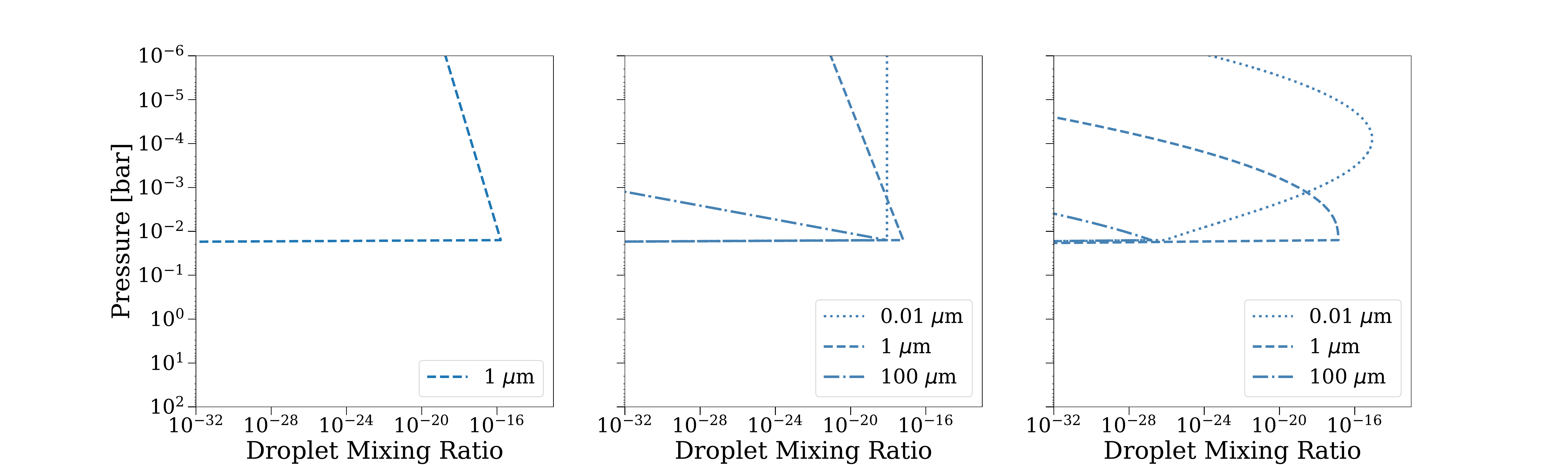}
\caption{Representative droplet \text{mixing ratio} profiles modeled in the one-size cloud model (left), the integrated cloud model (middle), and the A \& M cloud model (right, converted from condensate mixing ratios). For the latter two models, only profiles of three sizes of cloud droplets are shown. 61 bins of particle sizes from 0.001 to 1000 $\mu$m are included in the actual modeling.  They primarily differ in the shape of their vertical profiles. \label{fig:cloudmodels}}
\end{figure*}

Fig. \ref{fig:cloudmodels} illustrates the differences in the vertical droplet distributions under the different Mie scattering cloud parameterizations.

\subsubsection{Parameterized Haze Model for Warm-Neptunes} \label{subsubsec:hazemodel}
Clouds and hazes in the atmospheres of warm-Neptunes are likely to differ from those in hot Jupiter's due to the vastly different chemistry resulting from cooler temperatures ($\lesssim$ 1000 K). The presence of CH$_4$ and NH$_{3}$ provide the ingredients necessary to facilitate the photochemical production of upper atmospheric hydrocarbon/nitrile hazes (e.g. \citealt{Yung1984, Moses2011, Miller-Ricci Kempton2012, Morley2013, Adams2019}). While condensate clouds such as sulfide and chlorine species (KCl, NaCl, ZnS, Na$_2$S, etc.) likely exist in atmospheres below 900 K \citep{Morley2012}, they have proven difficult to constrain, considering the observed flat transmission spectra of a number of warm-Neptune-like worlds (\citealt{Morley2015}). These condensate clouds are likely to be obscured by the presence of photochemical hazes. 

To parameterize a photochemical haze, 
we qualitatively followed the results
of \cite{Kawashima2018} -- 
that haze particles exist in a much broader region in the atmosphere with a variety of different sizes, as a consequence of particle growth and settling, motivating us to assume that the haze can exist in the deep atmosphere, rather than only high altitudes.
We modify the Integrated cloud model to describe the haze (hydrocarbon mixture of \textit{tholin} and \textit{hexene} (C$_6$H$_{12}$)) with four parameters: the haze mixing ratio at the haze base, the mean haze particle radius at the haze base, the haze base pressure and the effective tholin fraction (ranging from 0 to 1). 
The last parameter describes the tholin fraction in the mixture in terms of its contribution as an opacity source. Optical properties, again, come from \citet{Wakeford2015}.

\bigskip
Table \ref{tab:models} summarizes the parameterizations of the four cloud models for the hot-Jupiter and the haze model for the warm-Neptune described above.

\begin{table*}[!ht]
\centering
\caption{Parameterizations of Cloud/haze Models Adopted}
\label{tab:models}

\begin{tabular}{lll}
\hline
\hline

Model name                             &  Parameters   &      Description  \\ \hline 
\multirow{3}{*}{\vtop{\hbox{\strut The Power Law Haze + Gray}\hbox{\strut cloud Model}}} & $P_c$   &   The ad-hoc cloud top pressure \\    & $\sigma_0$   &  \vtop{\hbox{\strut The amplitude of the haze cross-section}\hbox{\strut relative to H$_2$ Rayleigh scattering at 0.4 $\mu$m}}    \\  &    $\beta$   &  The slope of Rayleigh scattering    \\    \hline  \multirow{4}{*}{\vtop{\hbox{\strut The One-size}\hbox{\strut Cloud Model}}}    &    $P_{base}$   &  The cloud base pressure   \\   &    $f_{drop}$  &   The droplet mixing ratio at the cloud base   \\    &    $r$   &   The single particle size  \\   &   $\alpha$  &  Vertical profile index  \\ \hline     \multirow{3}{*}{\vtop{\hbox{\strut The Integrated}\hbox{\strut Cloud Model}}}  & $P_{base}$   &   The cloud base pressure     \\  &    $f_{drop}$   &    The droplet mixing ratio at the cloud base   \\  &    $\bar{r}$  &  Mean particle size at the cloud base   \\   \hline      \multirow{4}{*}{\vtop{\hbox{\strut The A \& M}\hbox{\strut Cloud Model (fiducial)}}}    &   $P_{base}$   &   The cloud base pressure   \\   &  $f_{cond}$   &    The condensate mixing ratio at the cloud base    \\  &   $K_{zz}$  &   Eddy diffusion coefficient    \\  &   $f_{sed}$  &  \vtop{\hbox{\strut The ratio of sedimentation velocity to}\hbox{\strut characteristic vertical mixing velocity}}   \\  \hline      \multirow{4}{*}{\vtop{\hbox{\strut The Integrated}\hbox{\strut Haze Model}}}              & $P_{base}$    &   The haze base pressure       \\
                      &  $f_{h}$   &  The haze mixing ratio at the haze base       \\
    & $\bar{r}$   &   Mean particle size at the haze base \\
     &    $F_{tholin}$   &   Effective tholin fraction in haze mixture \\ 
                                       \hline
\end{tabular}%
\end{table*}

\subsection{Forward Model} \label{subsec:forward}
We modify the atmospheric transmission spectrum model\footnote{A similar publically available version at https://github.com/ExoCTK/chimera} developed in \cite{Kreidberg2015, Kreidberg2018} and \cite{Greene2016} to compute transmission spectra for an atmosphere given a temperature profile, composition, cloud properties and other relevant planetary system parameters (Table \ref{tab:values}). 
The model divides the planet into annuli and computes the integrated slant optical depth and transmittance along each tangent height then integrates the slant transmittance with tangent height \citep{Brown2001, Tinetti2012}. We include gas absorption from H$_2$O, CO$_2$, CO, CH$_4$, H$_2$-H$_2$/He CIA, NH$_3$, Na, K, TiO, VO, C$_2$H$_2$, HCN, H$_2$S, and FeH with the pre-tabulated absorption cross-sections from \cite{Freedman2008, Freedman2014} and \cite{Lupu2016}, implemented within the correlated-k framework \citep{Lacis1991}. Continuum extinction from H$_2$/He Rayleigh scattering and condensate/haze Mie scattering are added into the total limb transmittance once the gaseous transmittances are computed within each k-ordinate. 
The molecular/atomic abundances are assumed to be in thermochemical equilibrium along the temperature-pressure profile and are computed using the NASA Chemical Equilibrium with Applications (CEA) Model \citep{Gorden1996} given the atmospheric metallicity and carbon-to-oxygen ratio. We further include as free parameters a carbon species and nitrogen species quench pressure parameter to approximate the effects of disequilibrium chemistry due to vertical mixing.   As in past works we parameterize the temperature-pressure profile with the analytic 3-parameter model \citep{Guillot2010} that takes in the irradiated temperature and mean opacities at infrared and visible wavelengths. Table 2 summarizes the key parameters required to compute the spectrum in our model.  The resulting spectrum is computed at an R = 100 (based on the pre-computed correlated-k bin opacities).  

\begin{table}
\scriptsize
\centering
\caption{Fiducial Planetary System, Atmospheric \& Cloud/haze Parameters and Values}
\label{tab:values}
\begin{threeparttable}

\begin{tabularx}{.5\textwidth}{llll}
\hline
\hline
Parameter   &  Description   &    \vtop{\hbox{\strut WASP 62b}\hbox{\strut  (Hot-Jupiter)}}        & \vtop{\hbox{\strut GJ 1214b}\hbox{\strut (Warm-Neptune)}}       \\  \hline
R$_*$ (R$_{\odot}$)\tnote{a}   &  Star radius  &  1.28   &  0.21     \\
R$_p$ (R$_{\text{Jup}}$)\tnote{a}  &  Planet radius  &  1.39  &  0.24    \\
M$_p$ (M$_{\text{Jup}}$) \tnote{a} & Planet mass  &  0.562  & 0.02    \\
T$_{\text{irr}}$ (K) \tnote{b}  & \vtop{\hbox{\strut Irradiation}\hbox{\strut temperature}}  &  1430  &  650   \\
Kir \tnote{b}   & \vtop{\hbox{\strut TP profile gray}\hbox{\strut IR opacity}}     &  10$^{-1.5}$  &  10$^{-1.5}$  \\
g$_\text{1}$ \tnote{b}   & \vtop{\hbox{\strut Single channel}\hbox{\strut Vis/IR opacity}}  &    10$^{-1}$ &   10$^{-1}$   \\
$\mathrm{[Fe/H]}$  &  Metallicity    &   0.0   &  1.9     \\
C/O     &   C-to-O ratio  &   10$^{-0.26}$  &  10$^{-0.26}$  \\
PQ$_{\text{C}}$ (bar)  &  \vtop{\hbox{\strut Quench pressure}\hbox{\strut for carbon species}\hbox{\strut CH$_4$, CO and H$_2$O}}  & 10$^{-5}$  &    10$^{-5}$  \\
PQ$_{\text{N}}$ (bar) & \vtop{\hbox{\strut Quench pressure}\hbox{\strut for nitrogen species}\hbox{\strut NH$_3$ and N$_2$}}   &   10$^{-5}$  &    10$^{-5}$  \\
P$_{\text{base}}$ (bar) \tnote{c,d}  &  See Table \ref{tab:models}  &    10$^{-1.8}$  &   10$^{1.5}$    \\
f$_{\text{cond}}$ \tnote{c}  &  See Table \ref{tab:models}   &    10$^{-4.15}$   &    -    \\
K$_{\text{zz}}$ \tnote{c}    &  See Table \ref{tab:models}  &    10$^{8.3}$    &  -  \\
f$_{\text{sed}}$ \tnote{c}   & See Table \ref{tab:models}  &   2.0       &    -   \\
f$_{\text{h}}$ \tnote{d}     & See Table \ref{tab:models}  &   -     &  10$^{-10}$  \\
$\bar{r}$ ($\mu$m) \tnote{d}  & See Table \ref{tab:models}  &   -     &  1.0   \\
 F$_{\text{tholin}}$ \tnote{d} & See Table \ref{tab:models}  &  -     &  0.7   \\  \hline
\end{tabularx}
\begin{tablenotes}
\footnotesize
\item[a] Adopted from exoplanets.org.
\item[b] The 3-parameter analytic model for temperature profile \citep{Guillot2010}.
\item[c] The four cloud parameters for WASP 62b are from the A \& M cloud model.
\item[d] The four haze parameters for GJ 1214b are from the integrated haze model.
\end{tablenotes}%

\end{threeparttable}
\end{table}

We simulated transmission spectra of our modeled planets using three \textit{JWST} instrument modes: NIRISS, covering 0.6 - 2.5 $\mu$m\footnote{NIRISS can cover 0.6 - 2.8 $\mu$m, but we applied a cutoff at 2.5 $\mu$m to avoid spectral contamination.}; NIRCam, covering 2.5 - 5.0 $\mu$m; and MIRI, covering 5.0 - 11 $\mu$m. 
The broad wavelength range covered by these instruments will show any power law slopes in the visible and near-IR range and resonance features at mid-IR, as well as distinguishable absorption features of major carbon-, oxygen- and nitrogen-bearing species in the atmospheres.
Compared to only using NIRSpec for 0.6 to 5 $\mu$m, the combination of NIRISS and NIRCam has the advantages of allowing slitless operation, larger flux limits and finer sampling of spatial features.
The details of the optics, spatial resolution and sampling of these instrument modes can be found in Table 4 and section 4 in \cite{Greene2016}.

We have simulated the noise (error bars) for all transmission spectra based on the model described in \cite{Greene2016}. 
For simplicity we assume purely photon noise limited observations for the hot-Jupiter WASP 62b.  This allows for us to readily re-scale the error bars for variable numbers of transits. 
For the generic warm-Neptune GJ 1214b, we simply scaled the noise for WASP 62b to 2.1 times itself.
Because the stellar brightness of WASP 62 and GJ 1214 are similar (especially, both H-, K-band magnitudes are close to 9), and that the transit duration of WASP 62b is about 4.4 times that of GJ 1214b\footnote{Based on data from exoplanets.org.},
resulting in a factor of 2.1 in the SNR of WASP 62b data relative to that of GJ 1214b.
Note that the different spectral types of WASP 62 and GJ 1214 lead to some magnitude differences at shorter wavelengths (e.g. J-band).
Considering we are only taking these planets as generic representatives of exoplanet types in this study, such differences are negligible and do not affect our conclusions.

All transmission spectra are binned to the final spatial resolution R = 100, except for the modes with R $\le$ 100, as a compromise between reaching the optimal signal-to-noise value and still resolving the major gas features. 
Note, as was done in  \cite{Feng2018} in order to mitigate the influence of random noise instances, we did not randomize the placement of data points for any simulated spectra.
This is because single random noise instance in the data can introduce unpredictable biases in the posteriors of retrieved parameters. As was pointed out in \cite{Caldas2019}, one needs to combine and average posterior distributions of retrievals obtained with different random noise instances to eliminate such biases. The larger number of noise instances, the smaller the biases we would expect from the posteriors.
\citet{Feng2018} performed such practice with merely 10 noise instances and has shown the average posteriors of retrievals from multiple random noise instances are close to the ones from non-randomized data, which can be indicated from the central limit theorem.

\subsubsection{Simulated \textit{JWST} Spectra} \label{subsubsec:noise}

\begin{figure*}[ht!]
\centering
\gridline{\fig{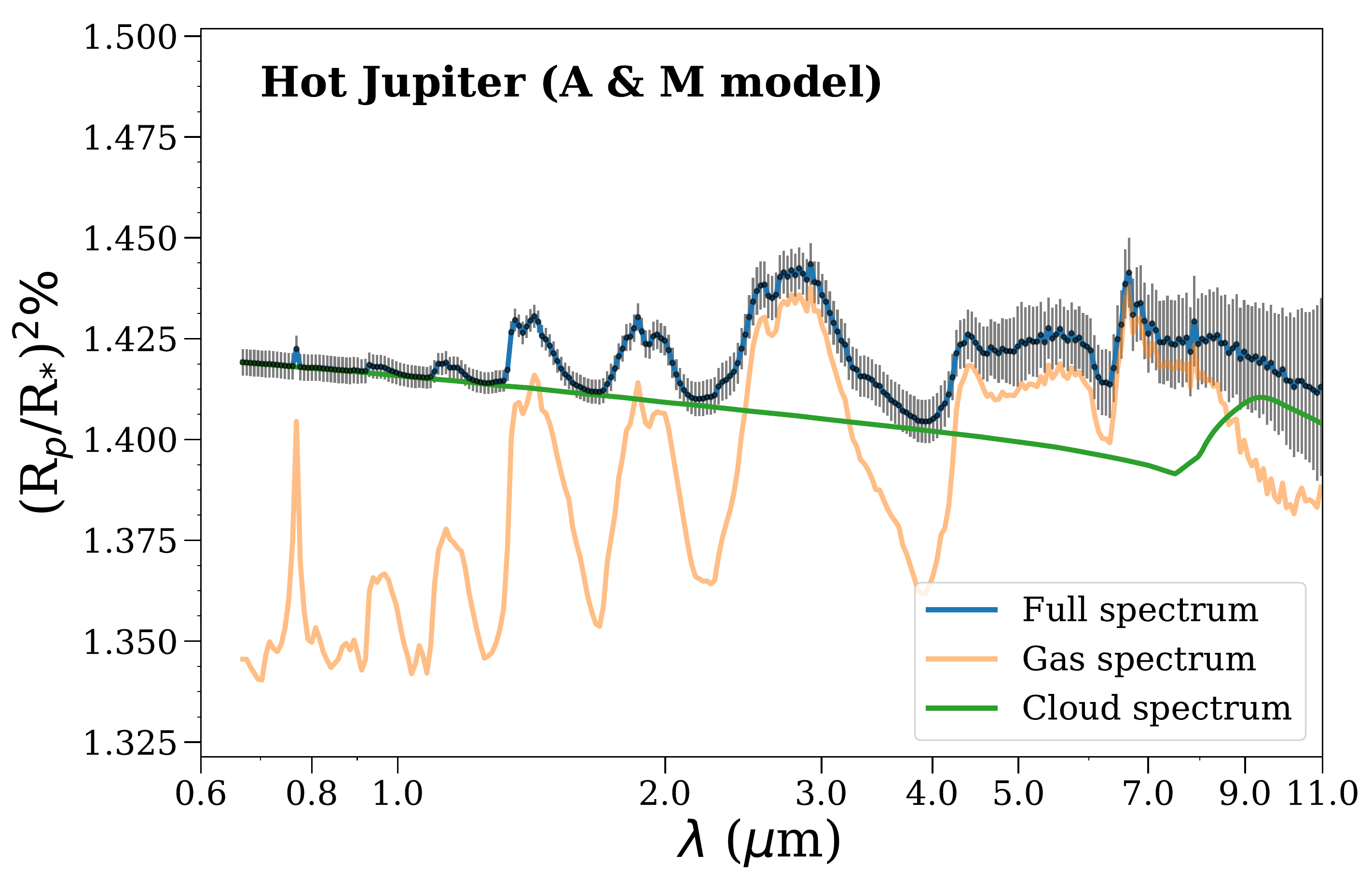}{0.5\textwidth}{(a)}
         \fig{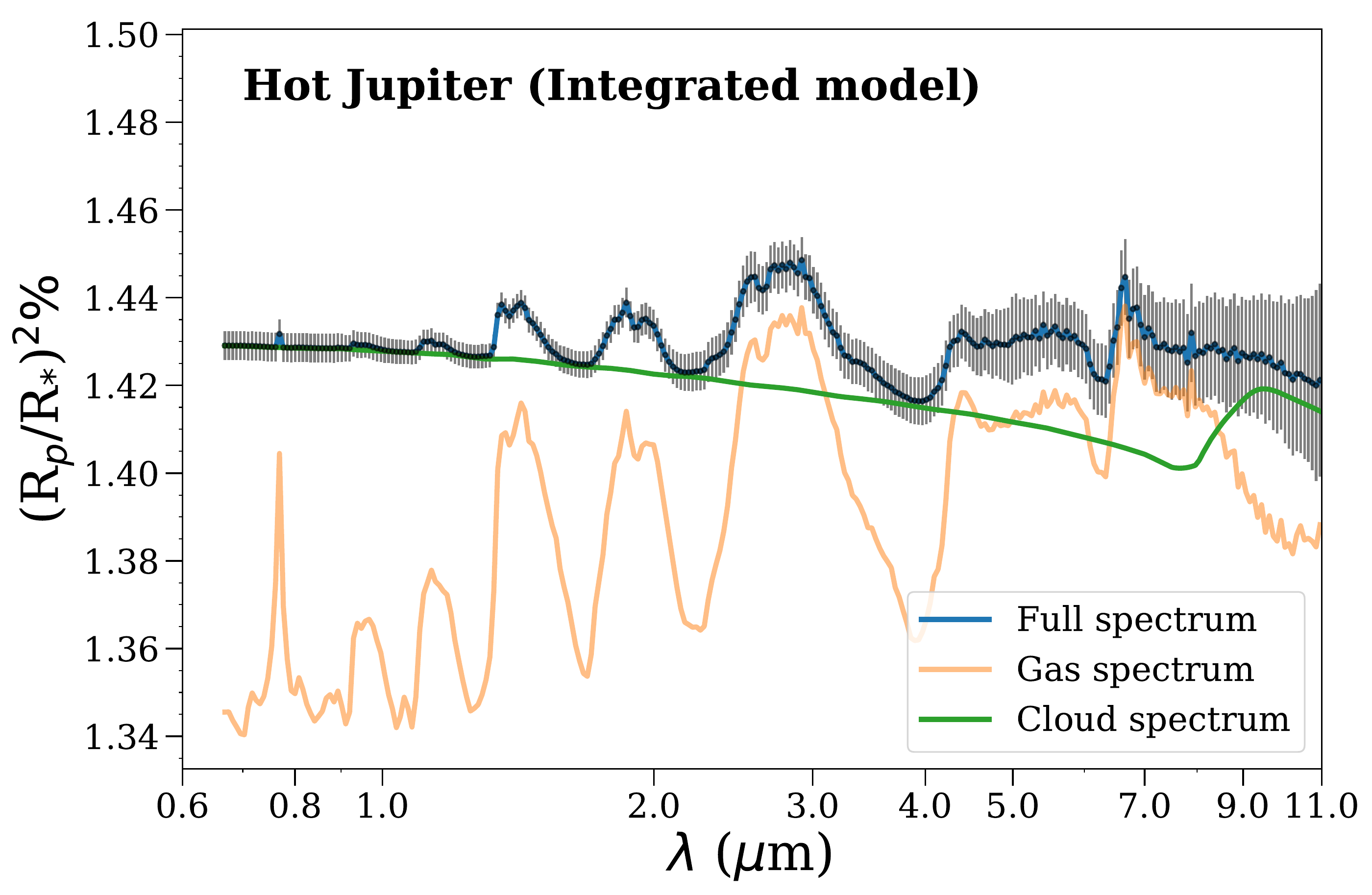}{0.5\textwidth}{(b)}
          }
\gridline{\fig{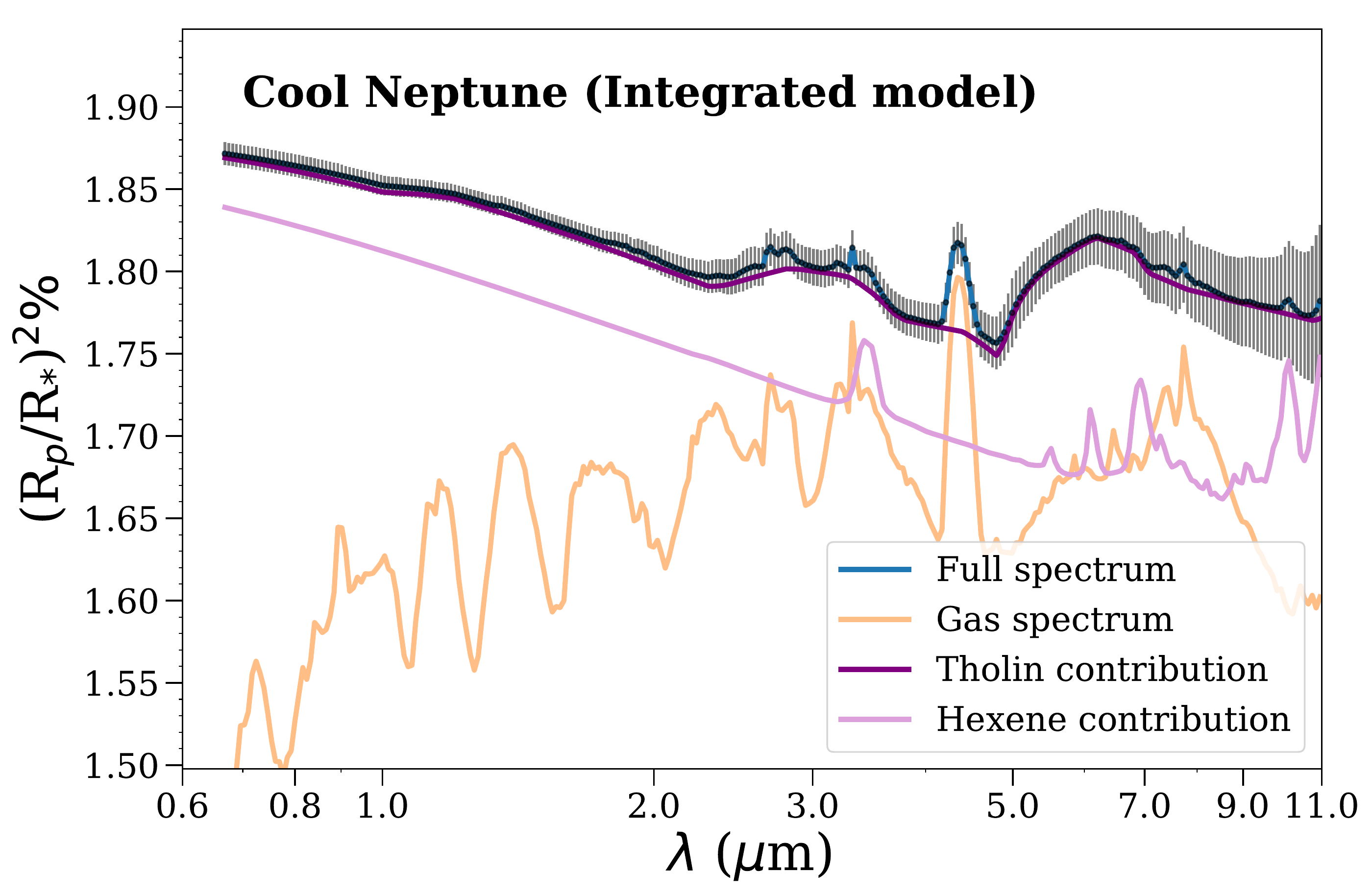}{0.5\textwidth}{(c)}
          \fig{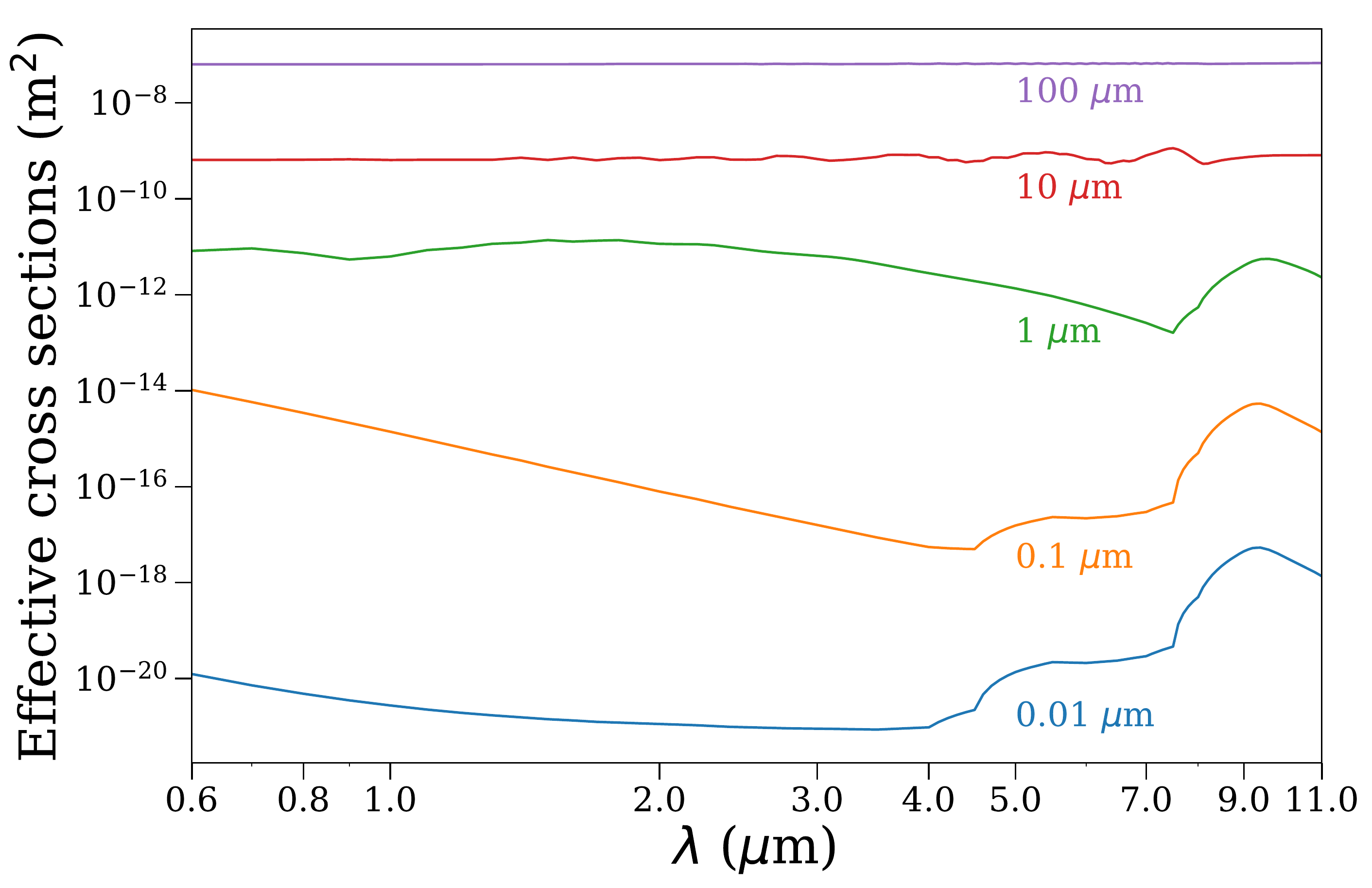}{0.5\textwidth}{(d)}
          }
\caption{Simulated transmission spectra for the hot Jupiter with MgSiO$_3$ clouds using the A \& M cloud model (a), using the integrated cloud model (b). 
The two cloud models produced cloud spectra similar to each other -- a gentle slope with a small resonance feature.
c: Simulated transmission spectrum for the cool Neptune with haze composed of tholin (70\%) and hexene (30\%), using the integrated haze model.
d: Effective cross sections of MgSiO$_3$ particles with different sizes, according to Mie theory. Please refer to \S~\ref{subsubsec:noise} in the text for a detailed description. \label{fig:specs}}
\end{figure*}

Figure \ref{fig:specs} shows the single-transit simulated transmission spectra for the hot-Jupiter and the warm-Neptune under several cloud model scenarios.  In the case of the hot-Jupiter (a \& b), the cloud introduces a gentle slope across the visible and near-IR range, and a small resonance feature from the Si-O bond in MgSiO$_3$ at 9 $\mu$m.

The A \& M cloud model and the integrated cloud model are able to closely replicate each other when given comparable parameter values, unsurprising given they both incorporate particle size distribution information. Panel (d) shows the effective cross sections of MgSiO$_3$ particles -- 
smaller droplets result in steeper slopes and larger resonance features, while larger droplets are nearly gray.
In both the A \& M cloud model and the integrated cloud model,
because various sizes of cloud droplets coexist in the atmosphere, neither the slope nor the resonance feature is rendered significant or negligible, thus the gentle slope and mid-IR feature. 

There has been much observational work reporting the existence of a steep slope in the visible range ($<0.8\ \mu$m.) of transmission spectra (e.g., HD 189733b), with explanations focusing on the presence of small high altitude particulates \citep{desEtangs2008, Pont2008, Pont2013, Lee2014, Wakeford2015}. The A \& M and integrated cloud models, where particles are brought up from some nominal condensate base pressure, are unable to qualitatively explain the observed slopes due to the relatively low abundance of small particles that make it to high altitudes.  Tuning these models to increase the abundance of small particles at high altitudes, also necessarily increases the presence of large particles that will wash out such slopes. That is, these quasi-self-consistent cloud parameterizations cannot reproduce the observed scattering slopes.   Only the power-law haze model and the single particle size model given an arbitrary droplet profile and small ($<0.1 \mu$m) particle size can reproduce currently observed slopes.


Figure \ref{fig:specs}c shows the simulated transmission spectrum for the warm-Neptune, with the effective tholin fraction at 70\%. 
The haze spectrum features a gentle and featureless slope at shorter wavelengths and multiple resonance features from the chemical bonds in tholin and hexene (e.g. the C-H bond) at longer wavelengths. 
As discussed in \S~ \ref{subsec:neptune}, these features should allow us to distinguish between more tholin-like hazes and pure hydrocarbon-like hazes. 
This synthetic spectrum is largely consistent with what was simulated with the photochemical model in \cite{Kawashima2018}, described in more detail in (\S~ \ref{subsec:neptune}).

The following sections \S~ \ref{subsubsec:control} to \S~ \ref{subsubsec:cloud_infer} will focus strictly on the hot-Jupiter scenario to address our major science questions.

\subsubsection{Froward Model Sensitivity Tests} \label{subsubsec:control}
\begin{figure*}[!ht]
\centering
\includegraphics[scale=0.27]{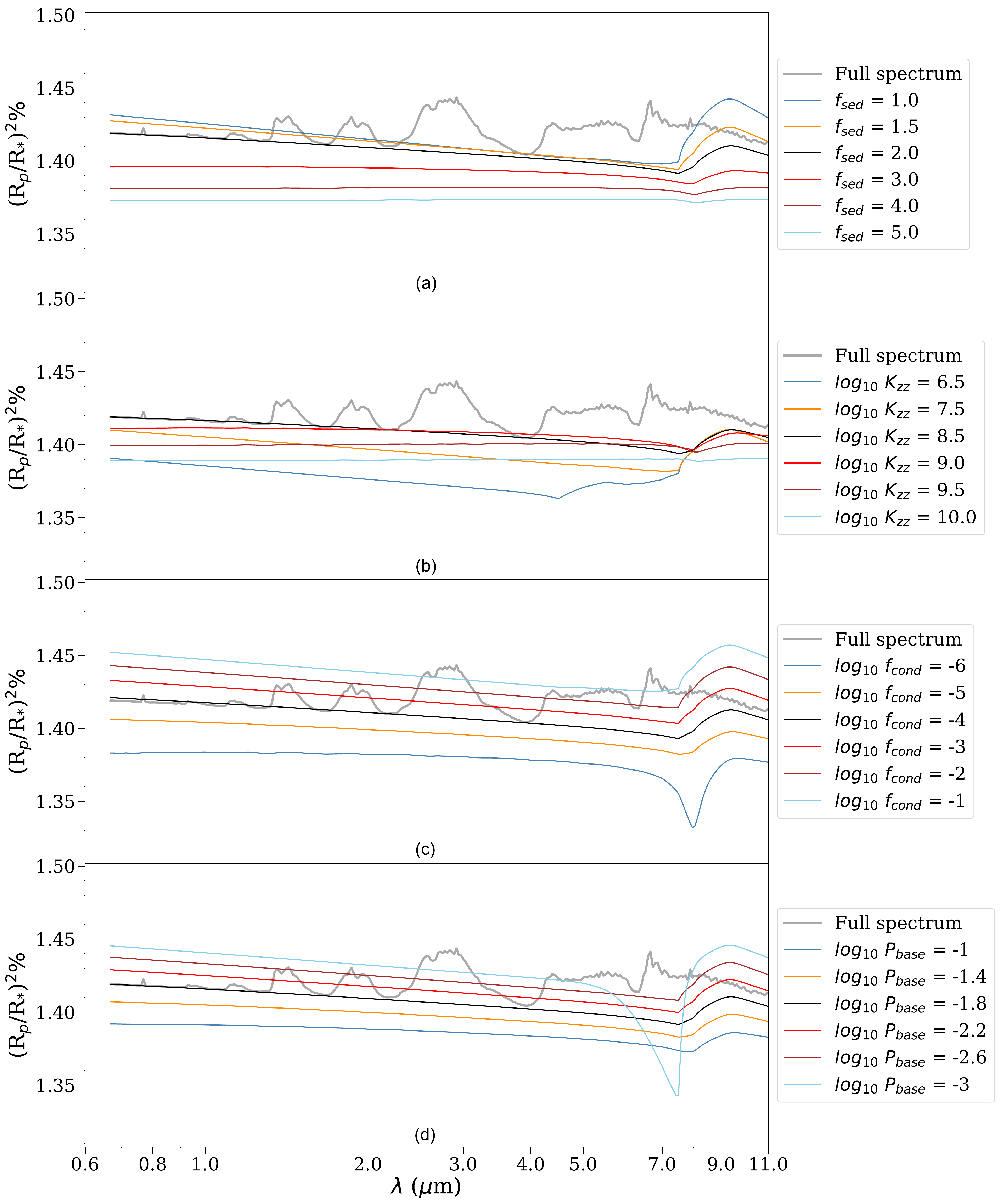}
\caption{Cloud parameters sensitivity tests.  The grey spectrum is the full hot-Jupiter spectrum from Figure \ref{fig:specs}a, while colored lines show the effects of changing the four cloud parameters in the A \& M model: (a) f$_{sed}$, (b) K$_{zz}$, (c) f$_{cond}$, and (d) P$_{base}$. 
See text for detailed discussion on how each parameter affects the cloud spectrum. 
\label{fig:control}}
\end{figure*}

Before performing any retrievals, we first attempt to build our intuition for how the various parameters of the A \& M model, our nominal cloud model, influence the cloud component of the observed spectrum.  Fig. \ref{fig:control} and items (a)-(d) summarize these effects. 

\begin{figure*}[ht!]
\centering
\hspace*{1.2cm}
\gridline{\fig{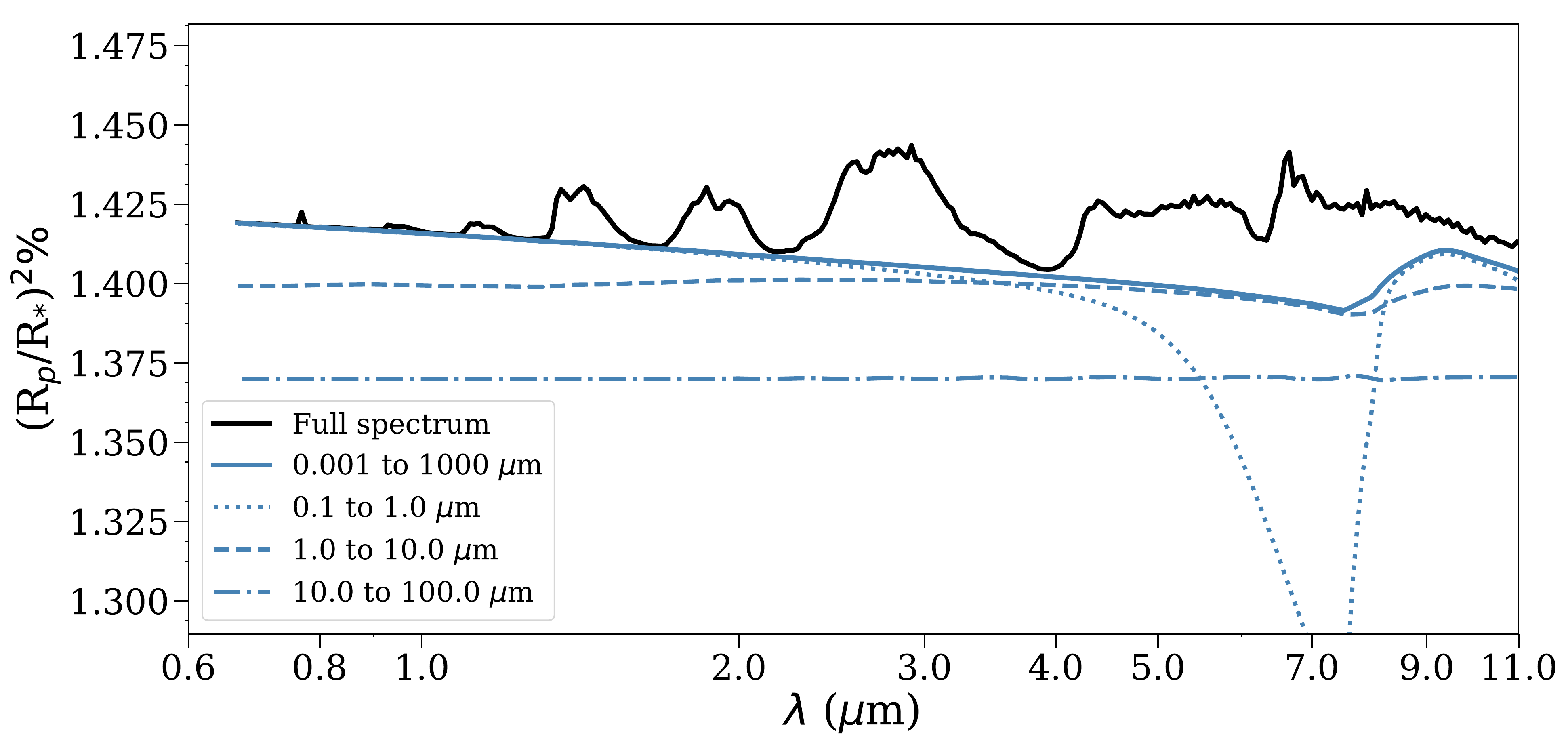}{0.55\textwidth}{(a)}
         \fig{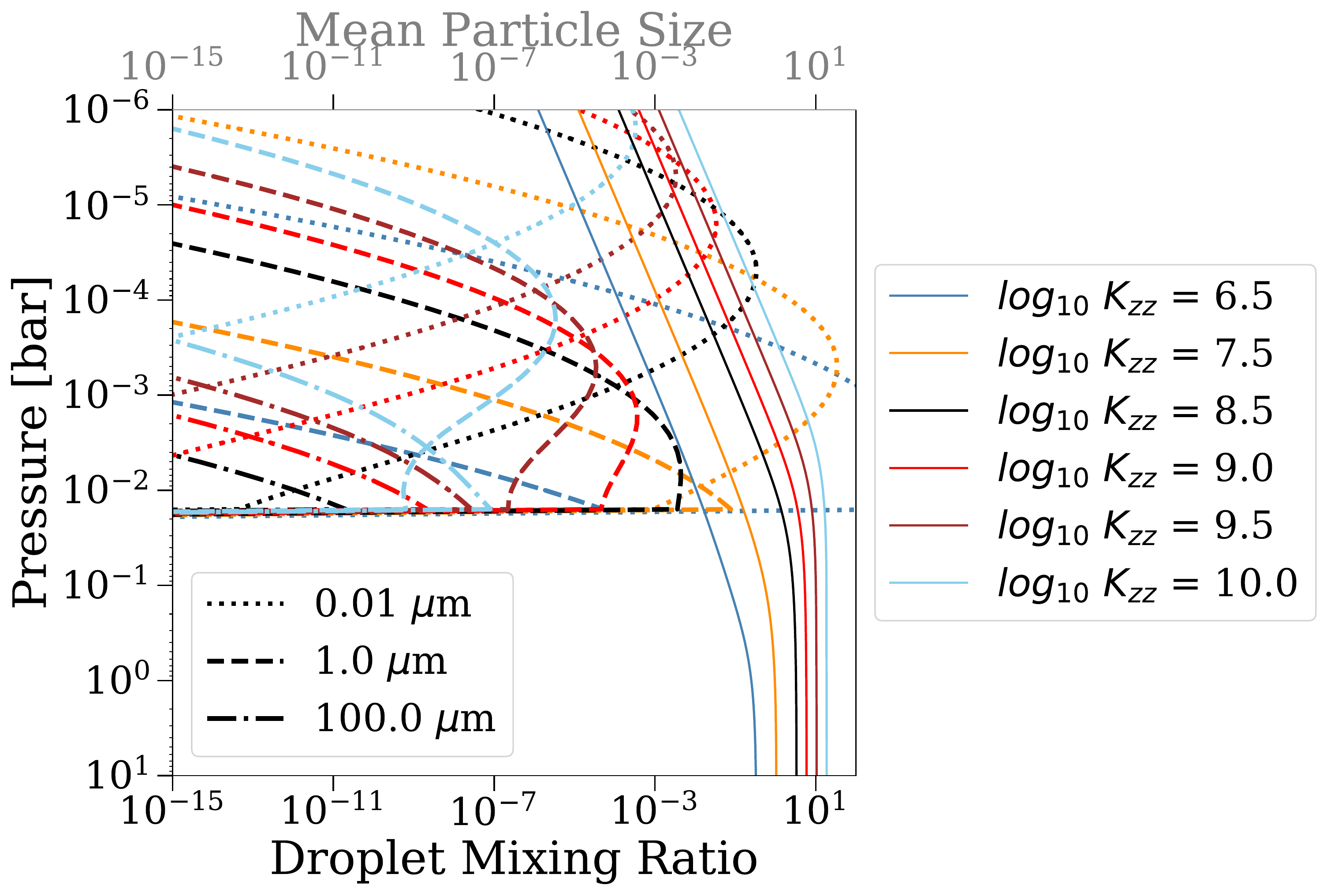}{0.45\textwidth}{(b)}
          }
\caption{a: The contributions from different particle size populations to the cloud spectrum. 0.001 to 0.1 $\mu$m and 100 to 1000 $\mu$m particles are not not shown because their contributions are negligibly small. The cloud spectrum is mainly shaped by particles within 0.1 to 10 $\mu$m -- the intermediate sizes. b: The vertical profiles of three representative sized particles, varying with $K_{zz}$. Dotted lines stand for the smallest size 0.01 $\mu$m, dashed lines are for intermediate size (1 $\mu$m) and dash-dot lines are for the largest (100 $\mu$m). The mean particle sizes are also shown as solid lines.  The ``cloud-base-effect'' is readily apparent in the 0.1 - 1 $\mu$ size bin as these sizes result in low enough optical depths to see the cloud base, until the Si-O resonance feature begins to dominate.  
\label{fig:contrib}}
\end{figure*}

\begin{enumerate}[label=(\alph*)]
\item $f_{sed}$: In the A \& M model, $f_{sed}$ controls 1) the compactness of clouds -- smaller $f_{sed}$ leads to more vertically extended clouds resulting in larger slant optical depths in the in upper atmosphere (depending on the location of the cloud base); 
and 2) the mean particle size of cloud droplets (as seen in Eqn. \ref{eq5}) -- a 
larger $f_{sed}$ results in larger sedimentation radii of cloud particles -- more compact clouds are composed of larger droplet sizes, which in turn washes away any slopes or cloud resonance-features (e.g., Fig. \ref{fig:specs}d).  The net effect of increasing $f_{sed}$  (Fig. \ref{fig:control}a) is to create a more compact, spectrally flat cloud. However, due to the compactness of the cloud, the ``flattening'' effect largely occurs at layers deeper than can be probed due the gas opacity alone.

\item $K_{zz}$: Larger $K_{zz}$ values are a proxy for stronger vertical mixing, resulting in larger mean particle sizes (via Eqn. \ref{eq5}) which should flatten the spectrum and increase the overall cloud opacity. This increased opacity (increasing the altitude at which the atmosphere becomes opaque) should ``shift'' the cloud spectrum upwards.  As shown in Fig. \ref{fig:control}b, we find this to be generally true up to moderate values of  $K_{zz}$ ($\sim 10^{8.5}$ or $10^9$) whereby the cloud opacity flattens and decreases, resulting in a downward shift in $(R_p/R_s)^2$.  This peculiar behaviour is related to the dominance of various particle size populations with altitude.  

Fig. \ref{fig:contrib}a shows the spectral contribution of different particle size bins (under the nominal model setup), which suggest that particles with intermediate sizes -- 0.1 to 10 $\mu$m -- shape the spectrum, thus the vertical distribution of these sized particles are what matters.
The vertical distribution of three representative sizes for each bin (0.01, 1 and 100 $\mu$m), is shown in Fig. \ref{fig:contrib}b. As $K_{zz}$ increases, small particles (e.g. 0.01 $\mu$m) are less abundant, and the population is shifted upwards; the abundance of large particles (100 $\mu$m) increases, and shifts to higher altitudes. However, the mixing ratio of intermediate-size particles (1 $\mu$m) is enhanced then reduced, with a turning point when $K_{zz}$ is about $10^{8.5}$ to $10^9$, consistent with what we see in Fig. \ref{fig:control}b. Overall, the mean particle size is still increasing with $K_{zz}$, consistent with Eqn. \ref{eq5}.


\item f$_{cond}$: 
The impact from f$_{cond}$ on the spectrum is straightforward. A lower abundance results in less opacity, resulting in a decreasing transit depth. If f$_{cond}$ is low enough, the cloud base becomes visible. The ``cloud base effect'' in the transmission spectrum, first proposed by \cite{Vahidinia2014}, is shown as an observable inflection point at some wavelength $\lambda_D$ in the spectrum. 
This is because there is more extinction above the cloud base (sensed by $\lambda < \lambda_D$) than beneath it, as both gas and the cloud contribute to the opacity. Light passing below the cloud base (sensed by $\lambda > \lambda_D$) experience less extinction as only molecular absorption is present. 
This results in a drop in the transmission spectrum. $\lambda_D$ is relevant to the abundance of clouds and the cloud base pressure (see detailed discussions in \cite{Vahidinia2014}). The observation of the cloud base effect can be used to break the degeneracy between f$_{cond}$ and P$_{base}$, which will be discussed in \S~ \ref{subsec:degen}.

\item P$_{base}$: 
 As other cloud parameters are unchanged, at each level the mean particle size (Eqn. \ref{eq5}) remains the same while the condensate mixing ratio at those level increases (as it is fixed at the cloud base). 
The particle population is re-partitioned accordingly with this change.
Thus, a higher cloud base results in a larger abundance of cloud particles associated with smaller sizes. The overall cloud opacity is increased, shifting the whole spectrum towards larger transit depths. 
When the cloud base reaches a certain high altitude and the population is dominated by sub-micron particles, the presence of the cloud base affect becomes apparent (near 7 $\mu$m in Fig. \ref{fig:control}d). 
Note that $\lambda_D$ and the depth of the drop appear different than in Fig. \ref{fig:control}c because the cloud abundance and the cloud base pressure are different.

\end{enumerate}

The above sensitivity test on how each cloud parameter affects the cloud spectrum will be useful when we explore correlations among these parameters (\S~ \ref{subsec:degen}).

\subsection{Parameter Estimation} \label{subsec:retrieval}
We use the \texttt{pymultinest} tool \citep{Buchner2014}, a python wrapper to the powerful \texttt{MultiNest}  nested sampling routine \citep{Skilling2004, Feroz2008}, to perform the parameter estimation and explore model degeneracies. Table \ref{tab:priors} lists the parameters and their corresponding prior ranges.

\begin{table*}[!ht]
\centering
\caption{The Retrieved Parameters and Prior Ranges for the fiducial Hot Jupiter,  A \& M Model scenario.}
\label{tab:priors}
\begin{threeparttable}
\begin{tabular*}{0.73\textwidth}{llll}
\hline
\hline

Parameter        &    Description        &   Input value    &   Prior range           \\ \hline
T$_{irr}$        &    Irradiation temperature (K)    &   1430           &   [300, 3000]           \\
log(Kir)         &    TP profile grey IR opacity (cm$^2$g$^{-1}$)         &    -1.5          &   [-3, 0]\tnote{a}        \\
log(g1)          &  Visible/IR opacity   &    -1            &   [-3, 1]       \\ 
$[\text{Fe/H}]$           &   Metallicty relative to solar          &   0.0            &   [-3,3]                \\ 
log(C/O)         &   C-to-O ratio        &    -0.26         &   [-2, 2]           \\
log(PQ$_{C}$)    &   Carbon quench pressure (bar)   &  -5          &   [-7.5, 1.5]      \\
log(PQ$_{N}$)    &   Nitrogen quench pressure (bar)   &  -5          &   [-7.5, 1.5]      \\
f$_{\text{sed}}$  &   \vtop{\hbox{\strut Ratio of sedimentation velocity to}\hbox{\strut  characteristic vertical mixing velocity}}                         &   2.0         &   [1.0, 5.0]\tnote{b}     \\
log(K$_{\text{zz}}$)   &  Eddy diffusion coefficient &    8.3     &  [6.5, 10.5]    \\
log(f$_{\text{cond}}$)  &  Condensate mixing ratio  &    -4.15    &   [-10.0, 0.0]  \\
log(P$_{\text{base}}$)  &  Cloud base pressure (bar)  &   -1.8  &  [-6.5, 1.5]  \\
xR$_p$            &     \vtop{\hbox{\strut Scaling to the fiducial 10 bar}\hbox{\strut  planet radius}}   & 1.0    & [0.5,1.5]   \\
                                       \hline
\end{tabular*}%
\begin{tablenotes}
\footnotesize
\item[a] \cite{Freedman2014}.
\item[b] \cite{Charnay2018, Ackerman2001}
\end{tablenotes}

\end{threeparttable}
\end{table*}

\subsubsection{Retrieval Self-validation} \label{subsubsec:valid}
In the retrieval self-validation test, the same model parameterization is used to create the simulated data as is used in the retrieval.  This is to ensure that there are no unforeseen biases, and to obtain intuition for the various degeneracies.  We use the  A \& M cloud model as the ``fiducial'' model with the nominal uncertainties on the data with all three modes (NIRISS, NIRCam and MIRI, see ``nominal'' case in Table \ref{tab:setup}).

\begin{figure*}[ht!]
\centering
\includegraphics[scale=0.27]{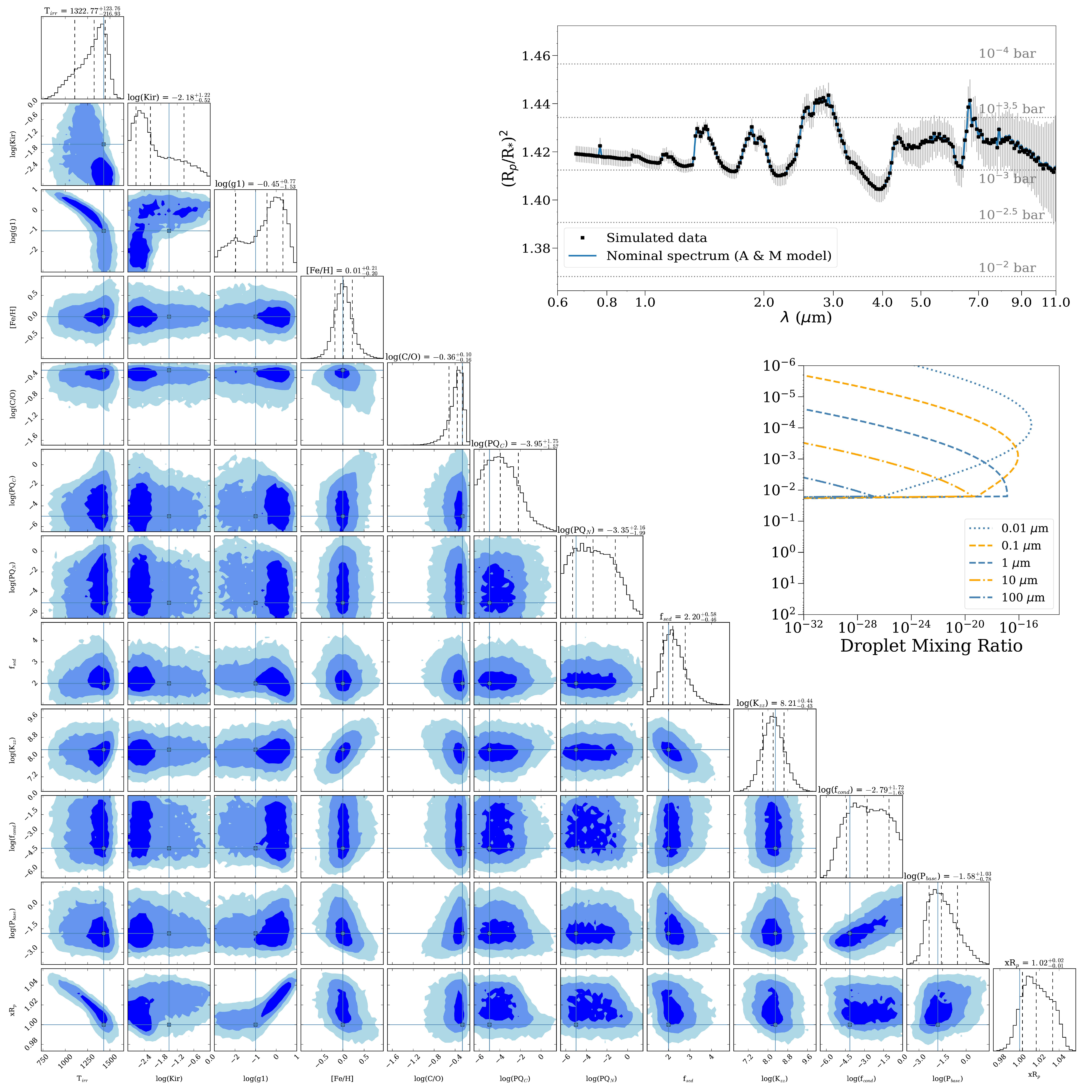}
\caption{Corner plot summarizing the posterior probability distributions under the nominal model/noise/instrumental setup. The blue solid line in each histogram represents the true value of the parameter in the forward model (Table \ref{tab:values}). The black dashed lines mark 16\%, 50\% and 84\% quantiles. On top of each histogram is the retrieved median value and $\pm 1-\sigma$ range. In between every two probability histograms are the two-dimensional 1-, 2- and 3-$\sigma$ contours illustrating the degeneracy between each parameter pair. Inset (a) presents the nominal model/simulated data. The horizontal dotted lines corrospond to the pressure levels probed by a given transit depth. Inset (b) shows the vertical distribution of different particle sizes in the fiducial A \& M model setup.  Most of the spectrum is probing the 0.3 - 1 mbar pressure range.  Relevant degeneracies are explained in detail in the text.
\label{fig:corner}}
\end{figure*}

Fig. \ref{fig:corner} summarizes the posterior probability distribution for the 12 retrieved parameters in this test (see Table \ref{tab:priors}). The parameters, T$_{irr}$, [M/H], C/O, f$_{sed}$, K$_{zz}$ and P$_{base}$, all have bounded uncertainties (e.g., they are constrained, not just upper or lower limits).

However, in this setup, we are unable to place a bounded constraint on f$_{cond}$ -- a lower limit only.  This is because the transmission spectrum only probes a fraction of the total cloud mass.  Decreasing f$_{cond}$ would allow us to see the cloud-base spectral signature, of which does not occur in this setup.  

In contrast to f$_{cond}$, P$_{base}$ can be constrained due to the increased sensitivity of the spectrum to subtle changes of this parameter (see Fig. \ref{fig:control}d). 
There exists a degeneracy (Fig. \ref{fig:condmixP}) between f$_{cond}$ and P$_{base}$ -- a deeper cloud can compensate for a larger condensate mixing ratio resulting in the same slant optical depth. 
We will discuss this degeneracy more in detail in \S~ \ref{subsubsec:degen_cP}.
It is interesting to point out that, if the cloud base is observable in the input transmission spectrum, 
the constraints on both f$_{cond}$ and P$_{base}$ can be more accurate, 
and that a different degeneracy between the two appears because the cloud base effect is quantitatively dependent on both parameters.

Another outstanding correlation among cloud parameters is the one between f$_{sed}$ and K$_{zz}$. Again, these degeneracies will be further explored in \S~ \ref{subsec:degen}.

\subsubsection{Experimental Set Up} \label{subsubsec:experim}
The key questions we ask in this work can be divided into: 
1) how do observation conditions (e.g. the noise level, the instrument modes) affect the inferred properties from transmission spectra? 
2) How do cloud parameterizations affect what we know about the atmosphere and the cloud?
We designed two sets of retrieval experiments to explore both questions accordingly.

In Experiment Set I, we set up the input simulated spectral data to explore the role of precision (SNR) and wavelength coverage. 
We adopted 8 different SNR levels by scaling the fiducial ``single transit error bars by factors of  $/2$, $/\sqrt[]{3}$, $/\sqrt[]{2}$, 1, $\sqrt[]{2}$, $\sqrt[]{3}$, $2$, $\sqrt[]{5}$, corresponding to either changing the number of transits, transit duration, or stellar magnitude. For the other half of this experiment, to explore the role of wavelength coverage, we selected one or two instrument modes among NIRISS, NIRCam and MIRI (See Table \ref{tab:setup} for the detailed instrument modes selection.) Note that \textit{JWST} instrument modes operate independently - e.g. obtaining the full spectrum from 0.6 - 11 $\mu$m requires observations of three separate transits \citep{Greene2016}. Although, a new mode has been proposed for NIRCam utilizing a Dispersed Hartmann Sensor (DHS) mode to enable simultaneous coverage of short wavelength (1 - 2 $\mu$m) with long wavelength (2.5 - 5 $\mu$m) \citep{Greene2016a, Schlawin2017, Schlawin2018}, though we did not explore that here.
In Experiment set I, retrieval forward models use the A \& M cloud model, the fiducial/baseline cloud model for the hot-Jupiter.

In Experiment Set II we start with the same nominal observational setup as above (A \& M cloud model and nominal noise), but applied different cloud parameterizations in the retrieval model.  

Table \ref{tab:setup} summarizes the set up for retrieval experiments described above.

\begin{table*}[!ht]
\centering
\caption{Summary of the observational and model scenarios explored in the Hot Jupiter scenario. The ``nominal'' case is used as a reference case. It serves as the ``true'' atmosphere when exploring other cloud parameterizations}
\label{tab:setup}

\begin{tabular}{lllll}
\hline
\hline
Purpose    &     \vtop{\hbox{\strut Error bar scaling}\hbox{\strut of input data}}   &  Instrument modes adopted     &      \vtop{\hbox{\strut Cloud model}\hbox{\strut for retrieval}}  &  Results \\ \hline
Nominal case    &  1     &  All modes (0.6 - 11.0 $\mu$m)   &  A \& M   &  Fig. \ref{fig:corner}  \\ \hline
\multirow{7}{*}{\vtop{\hbox{\strut Impact from}\hbox{\strut noise levels}}}   &  
1/2    &   All modes    &  A \& M   &  \multirow{7}{*}{Fig. \ref{fig:errbars}}   \\
       &  $1/\sqrt[]{3}$   &  All modes    &  A \& M   \\
       &  $1/\sqrt[]{2}$   &  All modes    &  A \& M   \\
       &  $\sqrt[]{2}$     &  All modes    &  A \& M   \\
       &  $\sqrt[]{3}$     &  All modes    &  A \& M   \\
       &  2                &  All modes    &  A \& M   \\
       &  $\sqrt[]{5}$     &  All modes    &  A \& M   \\ \hline
\multirow{6}{*}{\vtop{\hbox{\strut Impact from}\hbox{\strut instrument modes}}}   &
1      &  NIRISS (0.6 - 2.5 $\mu$m)  &   A \& M   & \multirow{6}{*}{Fig. \ref{fig:modes}}  \\
   &  1      &  NIRCam (2.5 - 5.0 $\mu$m)  &   A \& M   \\
   &  1      &  MIRI (5.0 - 11.0 $\mu$m)   &   A \& M  \\
   &  1      &  NIRISS + NIRCam (0.6 - 5.0 $\mu$m)  &    A \& M  \\
   &  1      &  NIRISS + MIRI (0.6 - 2.5, 5.0 - 11.0 $\mu$m)  &    A \& M   \\
   &  1      &  NIRCam + MIRI (2.5 - 11.0 $\mu$m)  &    A \& M   \\  \hline
\multirow{4}{*}{\vtop{\hbox{\strut Impact from cloud}\hbox{\strut parameterizarions}}}   &
1  &   All modes   &  Cloud-free  &  \multirow{4}{*}{Fig. \ref{fig:models_atm} \& \ref{fig:cloudparams}} \\
   &  1  &  All modes   &  Power law Haze + Gray  \\
   &  1  &  All modes   &  One-size  \\
   &  1  &  All modes   &  Integrated \\ \hline

\end{tabular}

\end{table*}

\section{Retrieval Results} \label{sec:results}
Fig. \ref{fig:errbars} to \ref{fig:cloudparams} summarize the marginal posterior distribution histograms of the 7 key parameters from every retrieval run in each of the above experiments. 

\subsection{Experiment Set I: Observation Conditions} \label{subsec:res_obs}
\subsubsection{The Impact From Noise Levels}

\begin{figure*}[ht!]
\centering
\includegraphics[scale=0.4]{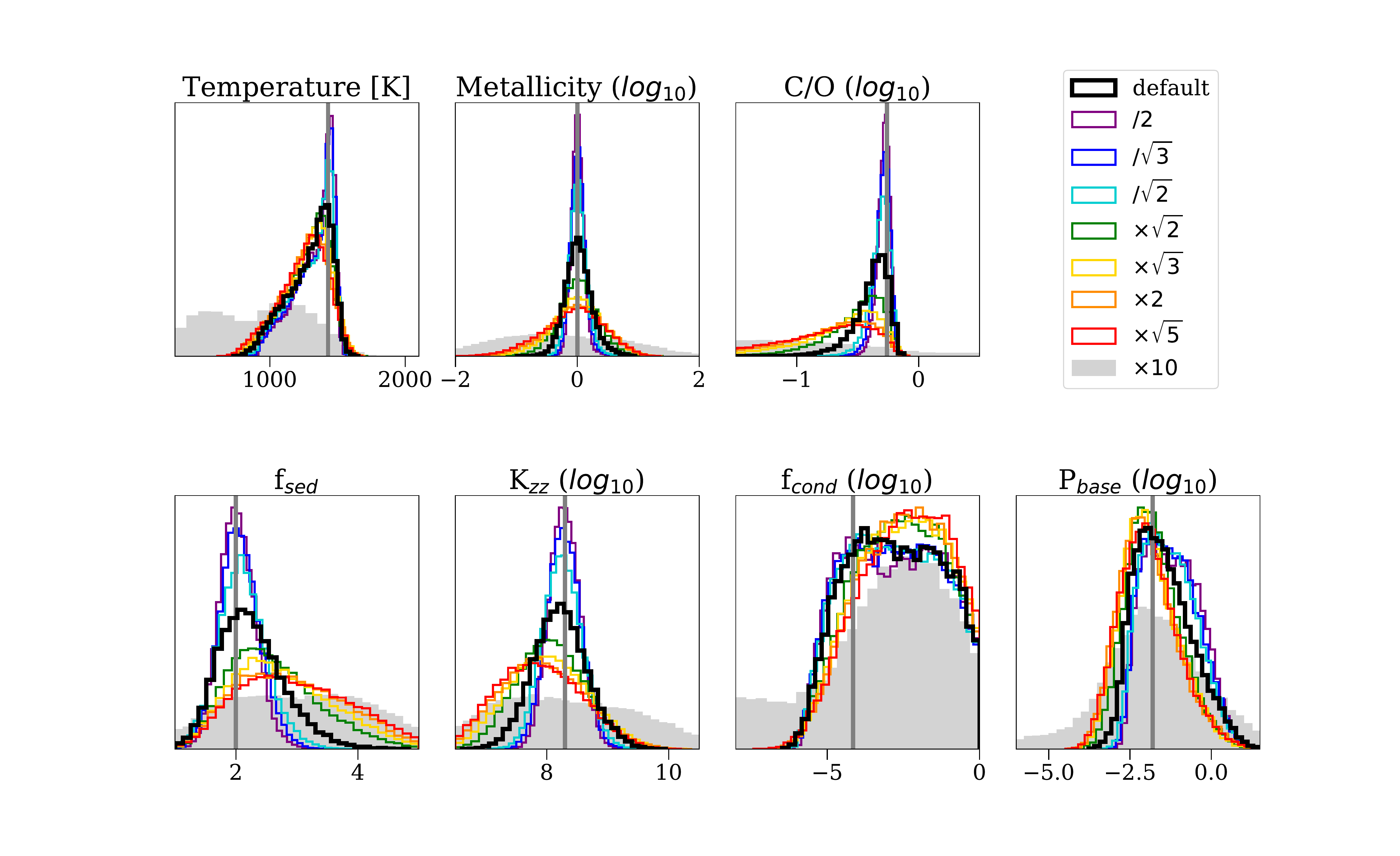}
\caption{Summary of the marginalized posterior probability distributions as a function of spectral precision given by a ``scaling'' to the default errors (black). The solid gray lines represent the truth values (Table \ref{tab:priors}). As expected, parameter constraints generally worsen when the spectral uncertainties grow larger.  \label{fig:errbars}}
\end{figure*}

Fig. \ref{fig:errbars} demonstrates the effect of spectral precision (e.g., fiducial error bar scaling) on each parameter. The retrieved uncertainties on the atmospheric properties increase as the transmission spectral precision gets larger.  For most key parameters, lower spectral precision not only increases parameter uncertainties, but also drives a ``shift'' in the parameter distributions due to the permittance of greater degeneracy.  For example lower preceision observations drive the retrieval towards a cooler atmosphere with a lower C/O ratio and a more compact, less abundant cloud located at slightly higher altitudes.

It is important to note that, even with the most precise data in this experiment (purple and blue colored histograms), we are still only able to obtain a lower limit on f$_{cond}$. Our lack of knowledge about the cloud abundance is intrinsic to the situation, regardless of precision, unless we are able to see the cloud base feature in the spectrum.

Another interesting find is that the noise level of the data does not significantly affect how well we can constrain the cloud base pressure. As pointed out in \S~ \ref{subsubsec:valid}, the cloud spectrum is highly sensitive to the cloud base pressure (Fig. \ref{fig:control}d). Over the range of uncertainties explored here, this sensitivity remained high, independent of SNR. However, when we scaled the uncertainties to 10 times their fiducial values (e.g., from 30 ppm to 300 ppm at 2 $\mu$m), the constraints started filling the prior range. Such a sensitivity is relevant to both the cloud base effect and the degeneracy between f$_{cond}$ and P$_{base}$.  The retrieval aims to maintain the ``slant optical depth'' over the atmospheric levels probed by the transit spectrum resulting in a degeneracy between P$_{base}$ and f$_{cond}$ -- a deeper cloud base can be compensated by an increased condensate base abundance, and vice-versa. Because of the spectral presence of the cloud base, the retrievals begin to disfavor base pressures $\sim <$ 10$^{-3}$ bar and low condensate abundances (low slant optical depths).  On the other end, the deeper base pressures are disfavored (larger than $\sim$1 bar) because f$_{cond}$, attempting to compensate for the drop in slant optical depth, would need to exceed 1. We will further elaborate on this effect in \S~ \ref{subsubsec:degen_cP} and Fig. \ref{fig:condmixPbase_err}.

\subsubsection{Impact From Instrument Modes}

\begin{figure*}[ht!]
\centering
\hspace*{-1.2cm}
\includegraphics[scale=0.4]{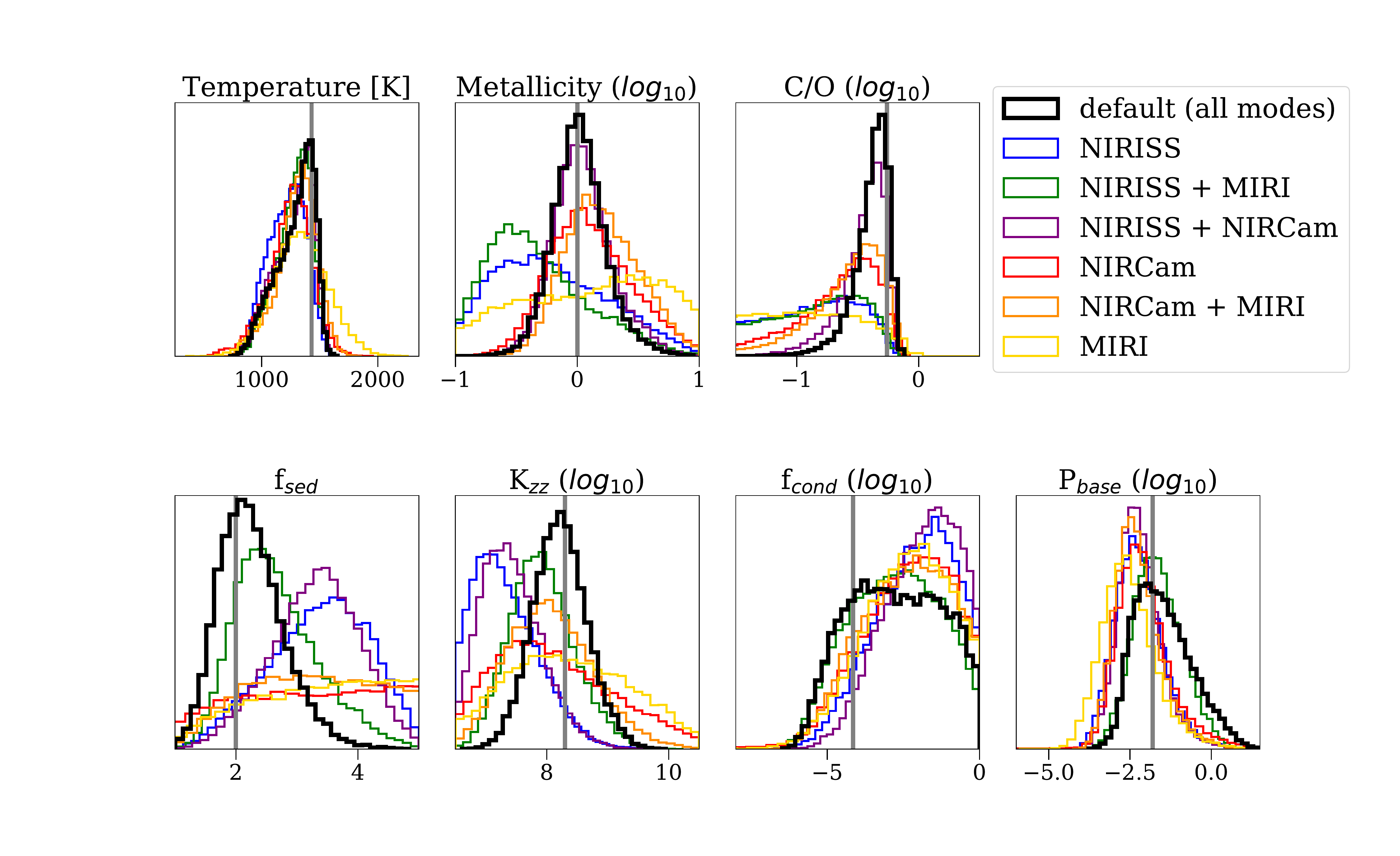}
\caption{Summary of the marginalized posterior probability distributions as a function of instrument mode combination. The solid gray lines represent the truth values (Table \ref{tab:priors}). At least two modes are required to obtain constraints comparable to those obtained with the full wavelength range (black).  The combination of NIRISS and MIRI is well suited for constraining the cloud properties (orange, bottom row) whereas the combination of NIRISS and NIRCam are suited for construing the atmospheric metallicity and carbon-to-oxygen ratio. \label{fig:modes}}
\end{figure*}

Fig. \ref{fig:modes} summarizes the constraints as a function of different combinations of instrument modes/wavelength coverage.  Here, we explore either one or two instrument modes, compared to using the full wavelength range.  In order to reasonably constrain the atmospheric temperature, metallicity, and C/O ratio, the NIRISS + NIRCam (0.6 - 5.0 $\mu$m, purple) combination is, at minimum, required. The wavelengths covered by NIRCam (3-5$\mu$m) in particular are critical for determining precision compositional constraints due to the presence of CO/CO$_2$ at these wavelengths (see also NIRcam + MIRI (2.5 - 11.0 $\mu$m, orange) and NIRCam-only (2.5 - 5.0 $\mu$m, red)). 
This is due to the presence of a strong CO feature at around 4.7 $\mu$m, which is highly sensitive to the metallicity and C/O ratio.  NIRISS + MIRI (0.6 - 2.5 and 5.0 - 11.0 $\mu$m, green) is the best combination for providing precision cloud property constraints. As expected, NIRISS is required to constrain the scattering slope at shorter wavelengths, while MIRI is important due to its coverage of the the mid-IR resonance features. Without MIRI observations ($\le$ 5.0 $\mu$m, blue and purple), a bias can occur where the cloud appears more vertically compact (higher f$_{sed}$) and composed of smaller particles (manifest with the lower K$_{zz}$) as the retrieval attempts to better fit the scattering slope, rather than the resonance feature. 
It is also worth noting that data from NIRISS alone ($\le$ 2.5 $\mu$m, blue) struggles to strongly constrain the metallicity and C/O ratio, when compared to combining it with additional modes suggesting that we need to go beyond 2.5 $\mu$m to characterize cloudy atmospheres more precisely.

To summarize, if the goal is to simply constrain atmospheric composition, then any combination of modes that includes NIRCam (or similar wavelength coverage modes like NIRSpec G395) is adequate; however if the science questions are driven by the desire to constrain cloud properties (at least in the context of the A\&M model) then at minimum, the combination of NIRISS + MIRI is needed. 


\subsection{Experiment Set II: Cloud Assumptions} \label{subsec:res_cloud}

\subsubsection{The Impact on Temperature, C/O, \& [Fe/H]}  \label{subsubsec:atm_infer}

\begin{figure*}[ht!]
\centering
\includegraphics[scale=0.27]{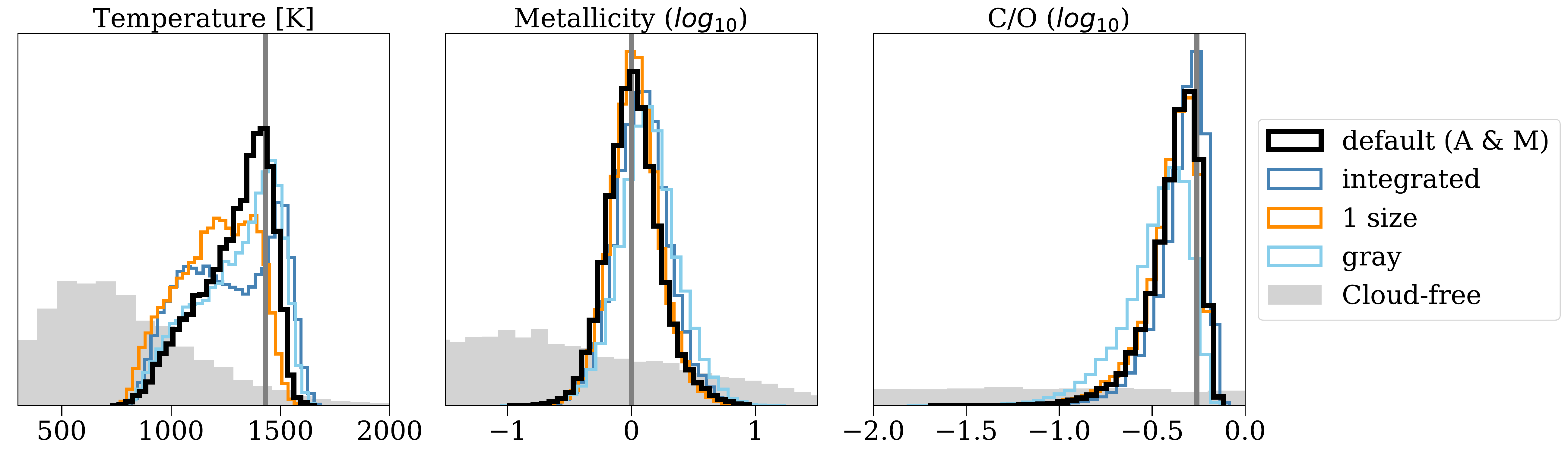}
\caption{Summary of the marginalized posterior probability distributions for temperature, metallicity, and C/O under different cloud modeling assumptions, where the underlying spectrum was produced with the A\&M parameterization. The truth values are given by the gray vertical lines (\ref{tab:priors}).  In general, as long as a cloud parameterization is included, these retrieved parameters are largely uninfluenced.   \label{fig:models_atm}}
\end{figure*}

The goal of Experiment II is to explore the impact on the retrieved composition under five different cloud assumptions/parameterizations in the retrieval: cloud-free, power law haze + gray model, one-size cloud model, integrated cloud model, and the A \& M cloud model. 


Fig. \ref{fig:models_atm} summarizes the constraints on the terminator temperature, metallicity, and C/O ratio under these different cloud modeling assumptions. 

As with numerous previous works (e.g. \citealt{Morley2013, Sing2016, Molliere2017}), we find that failing to include a cloud prescription can introduce large biases and vastly incorrect constraints (the gray shaded histograms).  In contrast, at least in this particular scenario, the four cloud models we tested all give similar and unbiased constraints on temperature and composition.

These relatively simple cloud prescriptions, though different in detail, appear to be effective in mitigating biases in non-cloud atmospheric parameters -- composition and temperature.   This is because the resulting best fit spectra all agree with each other until the resonance feature near 10 $\mu$m (Fig. \ref{fig:bestfit}). At these wavelengths, MIRI, in this particular observational scenario, the uncertainties are quite large and therefore this region of the spectrum is largely insensitive to the choice of cloud model.  Needless to say, this may not always be true in the presence of higher precision MIRI data.  


\subsubsection{The Impact on Cloud Properties} \label{subsubsec:cloud_infer}

\begin{figure*}[ht!]
\centering
\includegraphics[scale=0.37]{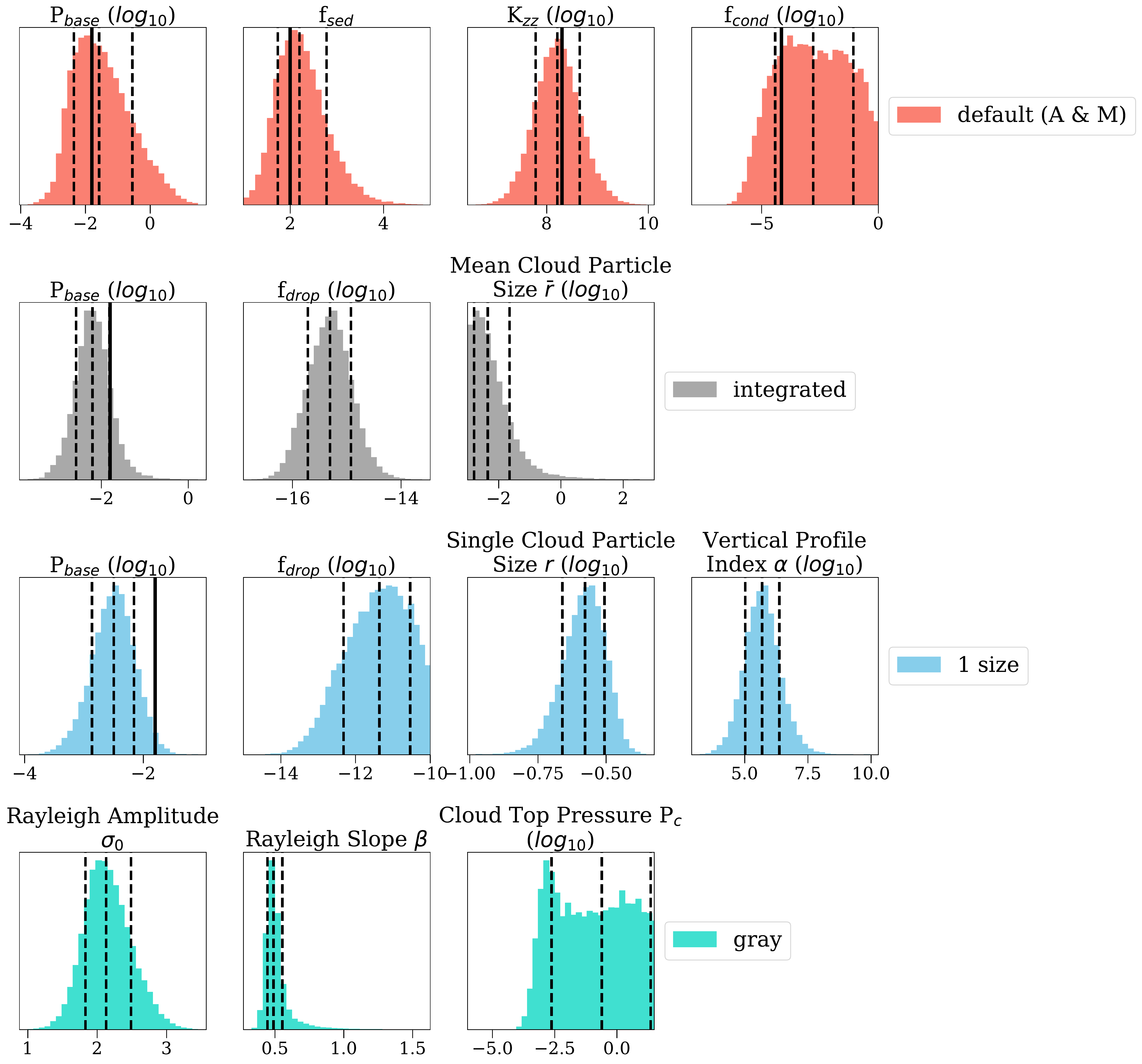}
\caption{Posterior probability distributions for the cloud parameters in the four different cloud models. The dashed black lines indicate the 16\%, 50\% and 84\% quantiles and the solid black lines denote the truth values (Table \ref{tab:priors}) where applicable (not all parameters have fiducial values).   \label{fig:cloudparams}}
\end{figure*}

It comes as no surprise that the choice of cloud model parameterization will result in different retrieved cloud properties.  Here, we explore the resulting constraints/biases within each of the different cloud parameterizations (integrated, one-size, and power law haze + gray cloud) assuming ``data'' generated with the fiducial A \& M cloud model. 

\begin{figure*}[ht!]
\centering
\includegraphics[scale=0.3]{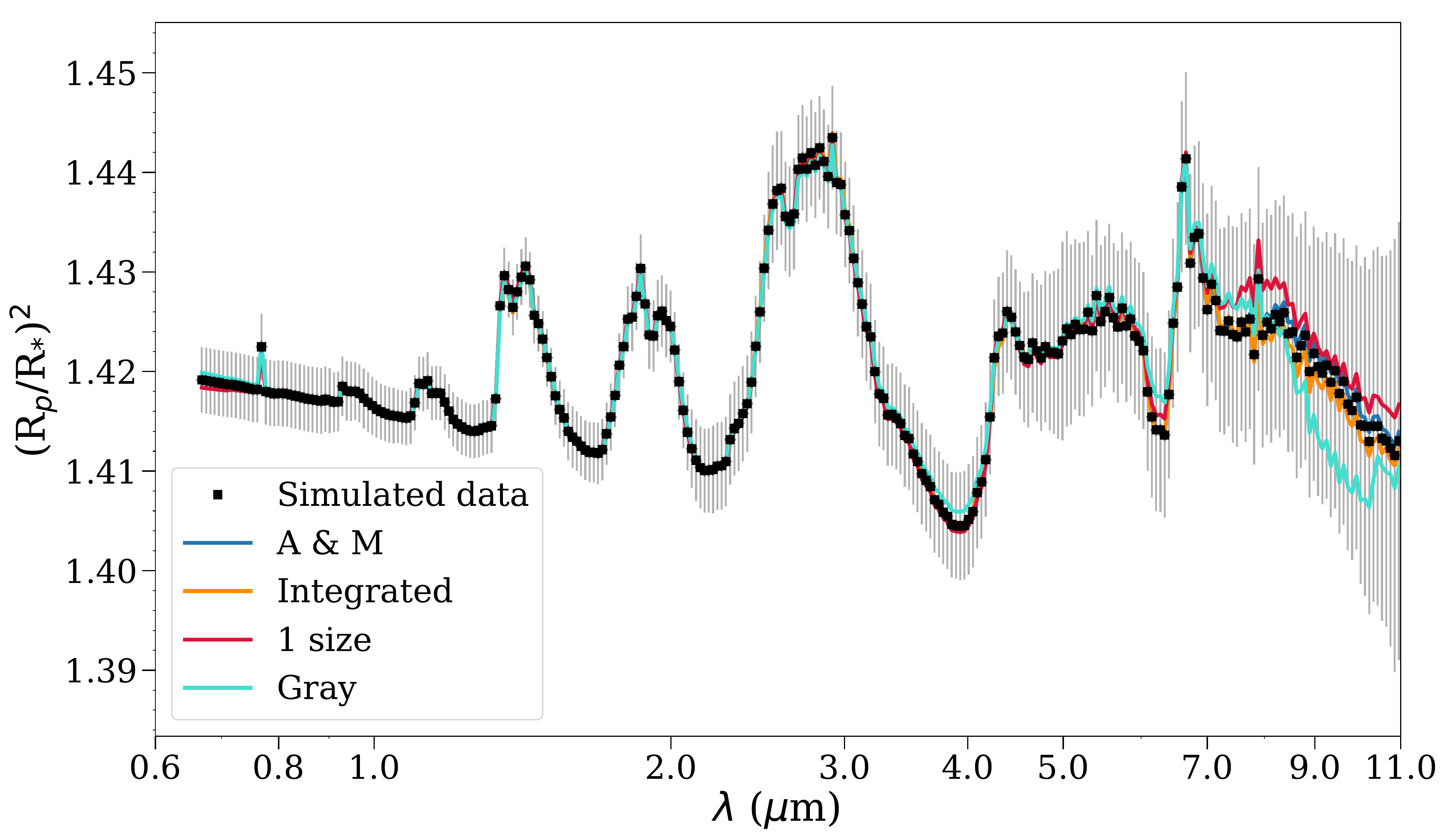}
\caption{Representative best-fit spectra under the different cloud models compared to the input simulated data.  All models are able to fit the simulated observations up until the resonance feature near 9$\mu$. \label{fig:bestfit}}
\end{figure*}

Figs. \ref{fig:cloudparams} and \ref{fig:bestfit} summarize, respectively, the constraints and representative best fits on the cloud properties under the four different cloud model assumptions.   While many of these models have similar parameters (e.g., the cloud base mixing ratio, the cloud base pressure), the parameterizations are different enough that they have really no direct relations.  Here we simply summarize the constraints one would get within each of these parameterizations. 
The purpose here is to show how different cloud models interpret the cloud properties from the same data set, and that not every parameterization gives a physically realistic picture of clouds.
The A \& M cloud constraints are shown in the top row of Fig. \ref{fig:cloudparams} -- same as what is in Fig. \ref{fig:corner}. 

The integrated cloud model (middle row, Fig. \ref{fig:cloudparams}), when retrieved upon a spectrum generated with the A \& M model, results in a very small modal particle size ($<$ 0.1 $\mu$m) with the cloud base located at a slightly lower pressure than the truth.  This is because of the differences in how the vertical particle sizes change within each parameterization, resulting in a greater sensitivity of the integrated cloud model to larger particles (see Fig. \ref{fig:cloudmodels}).  Thus the integrated cloud model tends to constrain a much smaller mean particle size -- defined at the cloud base -- to reduce the population of big particles aloft.  The resulting fit (Fig. \ref{fig:bestfit}, orange) is able to resonably match the true spectrum generated with the A \& M model. 
 

The one-size cloud model attempts to fit the spectrum with a very compact cloud (large $\alpha$ $\sim$ 5.6) with small particles ($\le$ 0.3 $\mu$m).  This would be considered a largely unphysical scenario as small particles are more readily lofted resulting in a more extended cloud (e.g., \citealt{Parmentier2013,Ackerman2001} ). This parameterzation also results in a larger retrieved f$_{drop}$ in order to achieve the required slant optical depths in the presence of smaller droplet sizes. As discussed in \S~ \ref{subsubsec:noise}, the shallow slope and the small mid-IR resonance feature produced in the fidcuial model are a result of different sized particle populations.  This morphology is impossible to reproduce with a single particle size as the small particles required to produce the near IR slope would result in a larger-than-acceptable mid-IR resonance feature (Fig. \ref{fig:bestfit}, pink).

The least physically motivated model, the power law haze + gray cloud model, retrieves a shallow power scattering slope ($\sim$ 0.5) and a cloud top pressure lower limit only.  The scattering slope index and the amplitude are able to mimic the gentle near IR slope produced by the small particle sizes in the A \& M model.   The ``cloud top-pressure'' concept is non-existent in the other models. This model simply tries to identify the pressure level at which the slant optical depths are large enough that the limb transmittance falls to zero.  The lower pressure level limit is a result of the unfavorable flattening of the spectrum if it were to move towards even lower pressures.  We will show that the upper limit of the cloud top pressure retrieved here matches the observable cloud pressure level in other cloud models when we revisit the discussion of the observable cloud mass in \S~ \ref{subsec:cloudmass}.  Finally, we note that this model is unable to produce a mid-IR resonance feature as that physics is simply not included in this parameterization (Fig. \ref{fig:bestfit}, cyan).

Again, each parameterization explored here carries different information regarding the cloud properties, that is, you retrieve what you parameterize.  Generally, all of the parameterizations are able to reproduce the optical-to-near IR slope. Only the droplet clouds are able to produce resonance features.  If the goal is to learn something about the cloud properties themselves, then the choice of cloud model matters significantly. Among the four cloud models, the physically motivated A \& M model is able to provide a self-consistent interpretation of clouds with only a few parameters and is therefore recommended for retrieving cloud properties.
However, if the goal is to simply determine other properties (like composition), at least in this example, the choice of cloud model seems to have very little affect.

\subsection{Notable Cloud Degeneracies} \label{subsec:degen}

Fig. \ref{fig:corner} shows several key correlations/degeneracies within the fiducial cloud model (A\&M).  Here we describe in more detail the sources of the most important degeneracies: that between f$_{sed}$ and K$_{zz}$ and between f$_{cond}$ and P$_{base}$.  Effectively, the retrieval tries to maintain the ``true'' sizes and abundance of particles by adjusting the condensate base abundance, $f_{sed}$, $K_{zz}$, and P$_{base}$.

\subsubsection{Degeneracy Between $f_{sed}$ and $K_{zz}$}   \label{subsubsec:fK}

\begin{figure*}[!ht]
\centering
\hspace*{-1.5cm}
\includegraphics[scale=0.19]{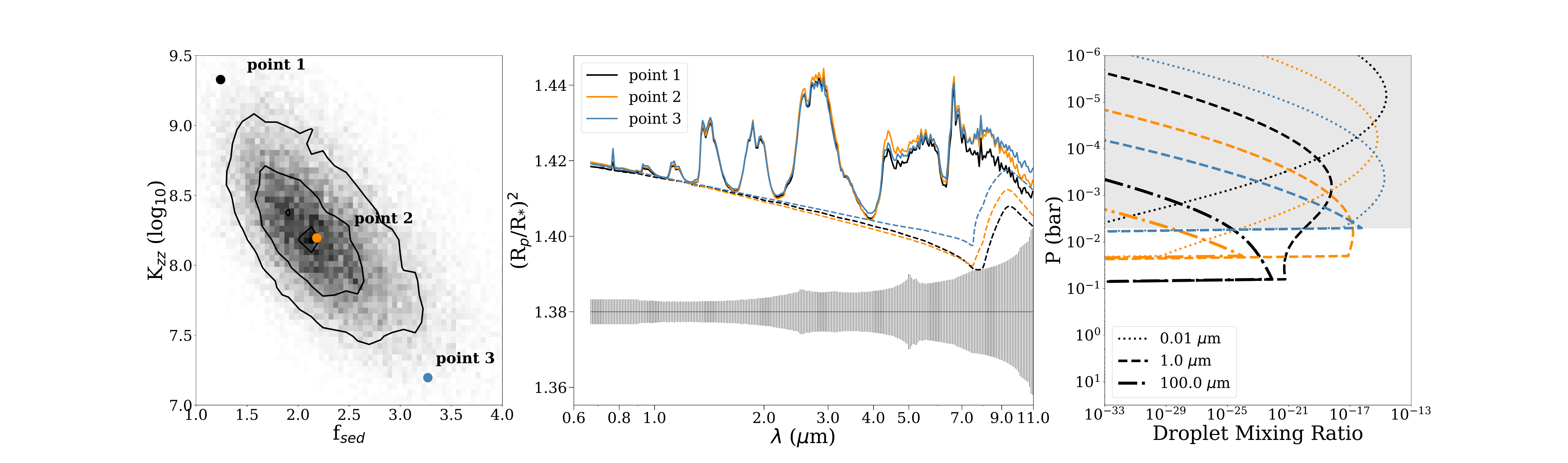}
\caption{The 2-D correlation between $f_{sed}$ and $K_{zz}$ (left panel) with corresponding analysis. Spectra (middle panel) and cloud droplet profiles (right most panel) are shown for three representative points (corresponding colors in each panel) along the correlation direction to provide insight into the nature of the degeneracy. The left panel shows the contours for the 68 and 95\% confidence intervals.  The middle panel shows the full transmission spectra for the 3 scenarios along with the opacity contribution from the condensate cloud (dashed spectra) for each case.  The ``grey'' envelope near the bottom represents the data error range with wavelength. The right panel shows the vertical mixing ratio profiles for select droplet sizes.  The light gray shaded area indicates the observable atmospheric pressure levels. See text for a detailed description. \label{fig:fsedKzz}}
\end{figure*}

Fig. \ref{fig:fsedKzz} summarizes the source of the degeneracy between $f_{sed}$ and $K_{zz}$. The left shows the marginalized 2-D probability distribution among these two parameters. We select three parameter combinations along the correlation direction to obtain insight on the behaviour of the degeneracy. Point 2 represents the maximum likelihood set of parameters (the other parameters appropriately adjusted), while points 1 and 3 are selected outside of the ``1-sigma'' contour.  The corresponding transmission spectra (cloud contribution shown as dashed lines) and cloud vertical profiles produced by these three sets of parameters are shown in the middle and right panels, respectively.  

We find that as $f_{sed}$ is increased (towards the right), $K_{zz}$ must decease (down).  This is unsurprising given that $f_{sed}$ and $K_{zz}$ are intrinsically correlated in the construction of the A \& M cloud model -- recall that $f_{sed}=v_{sed}/\omega^*$, and $\omega^*=K_{zz}/H$; so we have $f_{sed} \cdot K_{zz}=v_{sed} \cdot H$, where $v_{sed}$ is the sedimentation velocity of cloud droplets, $\omega^*$ is the characteristic vertical mixing velocity and $H$ is the atmospheric scale height.  Therefore, if other atmospheric conditions remain unchanged, $f_{sed}$ and $K_{zz}$ are inversely proportional.  Also recall that in Eqn. \ref{eq5}, these two parameters both control the mean particle size. Within the degeneracy, the anti-correlation of $f_{sed}$ and $K_{zz}$ arise to maintain the modal particle size at the pressure levels probed in the transmission spectrum. 

This degeneracy does not continue indefinitely. This is because $f_{sed}$ also directly controls the compactness of clouds. It is this additional feature of $f_{sed}$ that eventually breaks this degeneracy. From the right panel of Fig. \ref{fig:fsedKzz}, droplet sizes/mixing ratio, we can see that compared to point 2, point 1--black--(small $f_{sed}$, large $K_{zz}$) produces a highly vertically extended cloud with a base at deeper pressure level,
while point 3--blue--(large $f_{sed}$, small $K_{zz}$) produces a more compact cloud with a base at lower pressures (higher altitudes). 

The spectral consequences of these changes are apparent in the middle panel.  The shorter wavelengths, where higher precision is obtained, are largely unaffected by the parameter ranges spanned in the left panel.  There always remains a gentle slope due to the consistent abundance of intermediate sized (1 $\mu$m) particles. The variation in $f_{sed}/K_{zz}$ influences more strongly the vertical distribution of the smaller particles (0.01 $\mu$m). This in turn influences the Si-O resonance feature between 9 and 10 $\mu$m. However, due to the lower precision over MIRI, the size of the resonance feature is largely unconstrained.

\begin{figure*}[h!]
\centering
\includegraphics[scale=0.22]{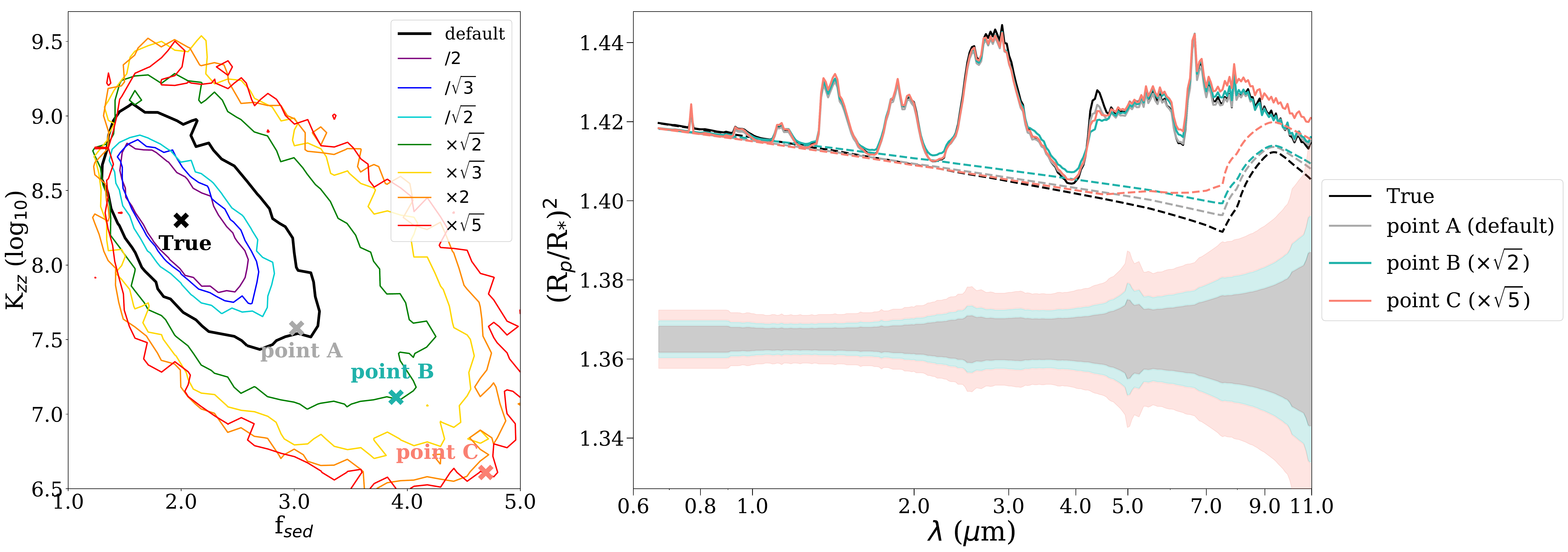}
\caption{Response of the $f_{sed}$ vs. K$_zz$ degeneracy (left) with spectral precision (right). The left panel shows the 68\% confidence intervals as a function of error bar scaling, with blue representing the higher precision data and red lower precision.  The ``x'' points along the correlation direction map to the spectra shown in the right panel.  The right panel shows the corresponding colored spectra (solid) and cloud contributions (dashed).  The colored shaded regions near the bottom represent the spectral uncertainty envelopes for the fiducial error bars and select scalings. ).
\label{fig:fsedKzz_err}}
\end{figure*}

The left panel of Fig. \ref{fig:fsedKzz_err} explores how increased signal-to-noise influences this degeneracy.  The degeneracy is largely ``broken'' from the high $f_{sed}$, low $K_{zz}$ end (bottom right). This is mostly an effect from our choice of the prior range of $f_{sed}$ (1.0 - 5.0, while the fiducial value is 2.0). We selected three points (parameter combinations) on the bottom right of the 16\% degeneracy contours at three noise levels (default, $\times \sqrt{2}$ and $\times \sqrt{5}$). We show the corresponding transmission spectra in the right panel of Fig. \ref{fig:fsedKzz_err}, with a best-fit spectrum at default SNR as a comparison (corresponds to the black cross marker). As expected, the other three spectra deviate from the best-fit spectrum , most noticeably at 0.6 - 1.0 $\mu$m, where the slope is present; at 4.0 - 5.5 $\mu$m, where the CO feature is located; and 8.0 - 11 $\mu$m, where the resonance feature rises. The model is less tolerable to such deviations when uncertainties decreases, hence the degeneracy is broken. However, further increasing the precision/SNR has little impact beyond a point simply because there is little change to the vertical distribution of particle sizes.


%

\subsubsection{Degeneracy Between f$_{cond}$ and P$_{base}$}  \label{subsubsec:degen_cP}

\begin{figure*}[!ht]
\centering
\hspace*{-1.5cm}
\includegraphics[scale=0.19]{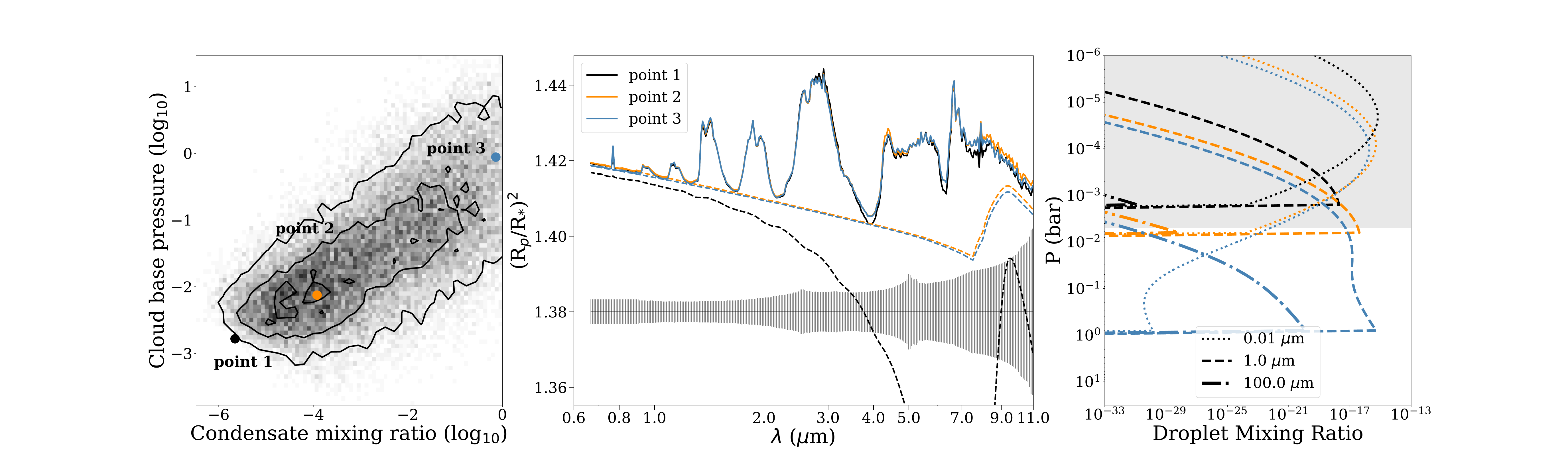}
\caption{The 2-D correlation between f$_{cond}$ and P$_{base}$ (left panel) with corresponding analysis. Similar to Figure \ref{fig:fsedKzz}, spectra (middle panel) and cloud droplet profiles (right most panel) are shown for three representative points (corresponding colors in each panel) along the correlation direction to provide insight into the nature of the degeneracy. The left panel shows the contours for the 68 and 95\% confidence intervals.  The middle panel shows the full transmission spectra for the 3 scenarios along with the opacity contribution from the condensate cloud (dashed spectra) for each case.  The ``grey'' envelope near the bottom represents the data error range with wavelength. The right panel shows the vertical mixing ratio profiles for select droplet sizes.  The light gray shaded area indicates the observable atmospheric pressure levels. See text for a detailed description. \label{fig:condmixP}}
\end{figure*}

f$_{cond}$ and P$_{base}$, like $f_{sed}$ and $K_{zz}$, are strongly correlated  (Fig. \ref{fig:condmixP} left panel) -- the larger the condensate abundance at cloud base, the deeper the cloud base. This is a fairly intuitive and unsurprising degeneracy.  For a given condensate vertical distribution, as the cloud slides towards ``deeper'' layers (higher pressures) the mixing ratio of the condensate must increase in order to maintain the same droplet mixing ratio at the observed altitudes.  For example, point 3 (in blue) in the left panel of Fig. \ref{fig:condmixP},  corresponds to a very deep cloud base with a large f$_{cond}$. At the altitudes probed by the transmission spectrum (Fig. \ref{fig:condmixP}, right panel -- light gray shaded area), this deep base-high condensate mixing ratio cloud (shown in blue) displays a similar profile as the cloud corresponding to point 2, shown in orange.  The corresponding spectra (middle panel) from parameter combinations point 2 and 3 are nearly indistinguishable for this reason.

Interestingly, along the correlation direction towards a lower P$_{base}$ and lower f$_{cond}$, an abrupt edge is encountered, but the degeneracy continues indefinitely in the opposite direction due to the aforementioned reasoning.  This abrupt discontinuation is caused by the ``cloud base effect'' \citep{Vahidinia2014}. At this location, point 1, the cloud base resides at a low enough pressure that the gas optical depth is small enough to see ``below'' the cloud base.  The corresponding cloud spectrum (black dashed curve, middle panel) shows an abrupt fall off near 3$\mu$m.  This drop in cloud opacity effectively results in a ``clear'' atmosphere situation from 4 - 11 $\mu$m -- so much so that data of the nominal precision are sufficient to rule out this scenario.

\begin{figure*}[h!]
\centering
\includegraphics[scale=0.22]{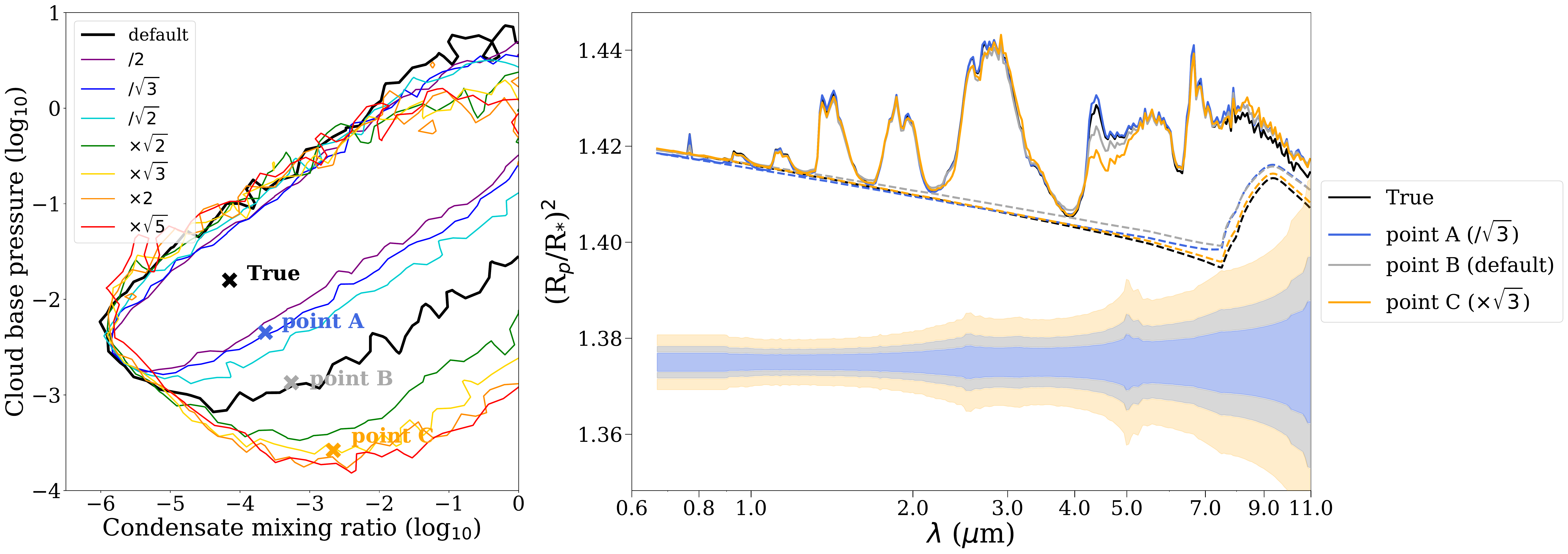}
\caption{Response of f$_{cond}$ vs. P$_{base}$ degeneracy (left) with spectral precision (right). Similar to Figure \ref{fig:fsedKzz_err}, the left panel shows the 68\% confidence intervals as a function of error bar scaling, with blue representing the higher precision data and red lower precision.  The ``x'' points along the correlation direction map to the spectra shown in the right panel.  The right panel shows the corresponding colored spectra (solid) and cloud contributions (dashed).  The colored shaded regions near the bottom represent the spectral uncertainty envelopes for the fiducial error bars and select scalings.).
\label{fig:condmixPbase_err}}
\end{figure*}

The left panel of Fig. \ref{fig:condmixPbase_err} demonstrates the behaviour of the degeneracy with SNR. Higher precision seems to decrease the uncertainty in the direction perpendicular to the main correlation, but offers little help in alleviating the overall mixing-ratio vs. base pressure degeneracy. 
This is why, as mentioned in \ref{subsec:res_obs},that increasing spectral precision does little to boost the constraint on the cloud base pressure. 
The right panel shows the best-fit transmission spectrum, as well as spectra corresponding to the three selected points in the correlation space along the direction perpendicular to the degeneracy (left panel). 





\subsection{Total Cloud Mass, Observable Cloud Mass} \label{subsec:cloudmass}

As was shown in \S~ \ref{subsubsec:valid}, we can only obtain a {\it lower limit} on f$_{cond}$ because we are only seeing a fraction of the total cloud.  Constraining a total cloud mass, however, would be desirable, as one in theory could say, constrain refractory compositions (e.g., Mg, Si, etc.) and link those back to planet formation.  We'd like to address whether or not it is possible to constrain the total cloud mass, and if not, how much of it we are actually privy to.
We should keep in mind that 1) the vertical cloud profile parameterization will play a large role in dictating the total cloud mass, so the result is strongly model-dependent; and 2) in general, the cloud base is obscured from view due to gas opacity in the atmosphere, and therefore a lower limit on the mass is more likely than a tight constraint.

The goal here is to compare the retrieved total cloud column to that of the ``true'' cloud column for that model and to the ``observed'' portion of cloud column accessible by the transmission spectrum. To compute the cloud column mass we integrate over the atmospheric column from the cloud base to the top of the atmosphere with, 
\begin{equation}\label{eq:cloud_col}
  M_{cloud}=\int_{P_{base}}^0 f(P)\frac{dP}{g}\frac{\mu}{\bar{\mu}}
\end{equation}
where $f(P)$ is the condensate mixing ratio profile, $g$ is the gravitational acceleration, $\mu$ is the molecular weight of the condensate and $\bar{\mu}$ is the mean molecular weight of the atmosphere.   Integrating equation \ref{eq:cloud_col} under the A \& M enstatite cloud model (equation \ref{eq:am_profile}) with the ``true'' values given in Table \ref{tab:priors}, we obtain a total ``true'' cloud column of $0.32\ kg\ m^{-2}$.   

\begin{figure}[h!]
\centering
\hspace*{-1.2cm}
\gridline{\fig{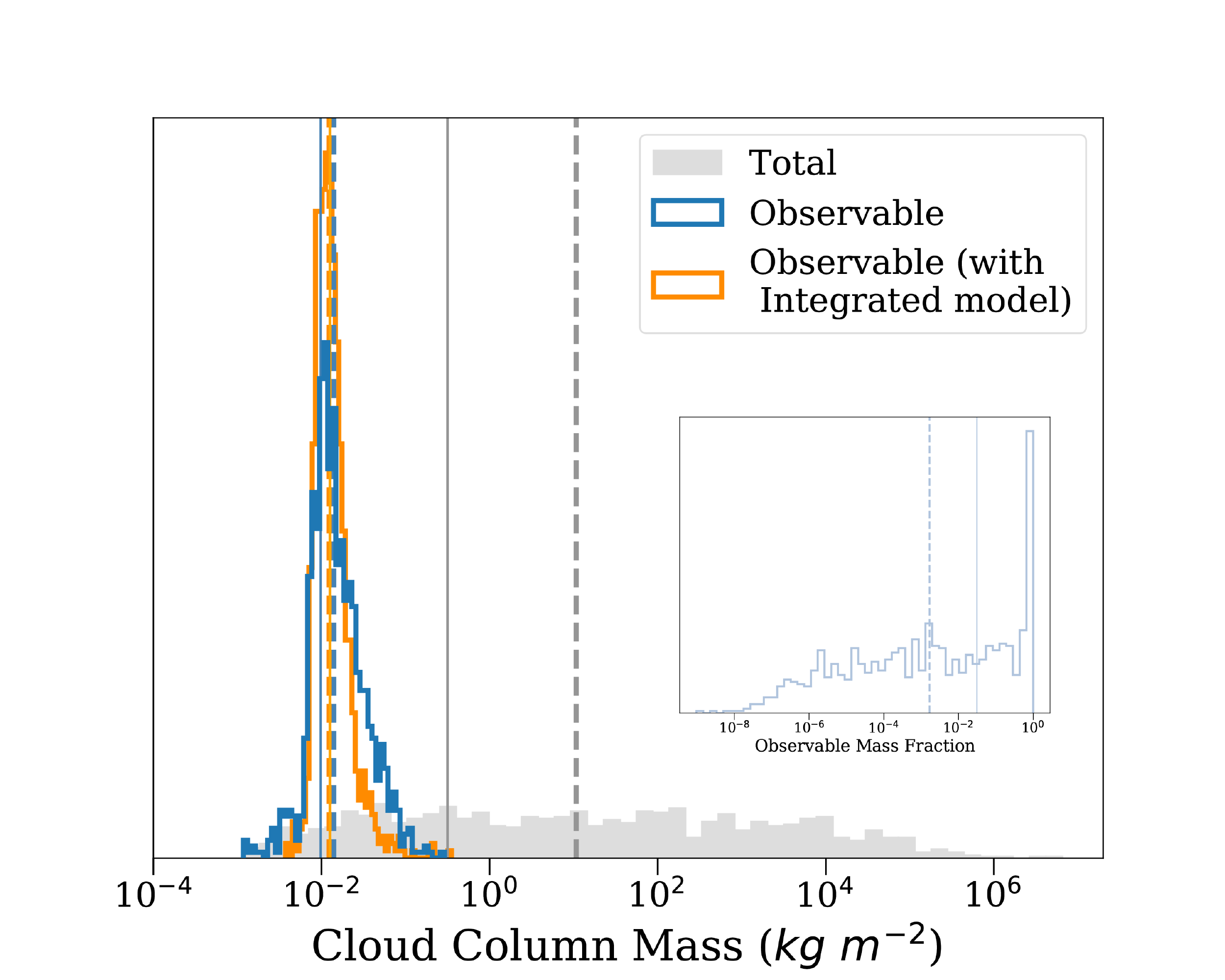}{0.5\textwidth}{}
          }
          \vspace{-5mm}
\caption{Constraints on the total cloud mass (solid gray histogram) and the observed cloud mass (blue histogram) under A \& M cloud parameterization. We also place the observed cloud mass (orange histogram) constrained with the integrated model as comparison. The cloud mass histograms were obtained by integrating equation \ref{eq:cloud_col} for a set of randomly drawn posterior samples.  The inset shows a histogram of the observed cloud {\it fraction}. The vertical solid lines indicate the ``true'' values (computed from the fiducial model from Table \ref{tab:values} and the dashed lines indicate the distribution medians). See text for details. \label{fig:mass_hist}}
\end{figure}

To compare the fraction of observable cloud mass to total cloud mass, we simply integrate equation \ref{eq:cloud_col} on randomly drawn posterior samples. The total cloud mass (shown as the solid grey histogram in Figure \ref{fig:mass_hist}) integrates from the base of the cloud to the top of the atmosphere.  The ``observable'' cloud mass (blue histogram in Figure \ref{fig:mass_hist}) integrates from the deepest pressure level at which the wavelength dependent limb transmittance is 0.5 for a particular posterior draw (approximately where the transit appears as an opaque disk).  
The integrated cloud model also constrains a similar observable cloud column mass (orange histogram) -- which is why these two models generate similar cloud spectra (Fig. \ref{fig:specs}a, b).
Under the A \& M cloud model we find that the ``deepest'' observable pressure occurs near log(P) $\sim$ -2.3 (recall the base is log(P) = -1.8). Upon integrating from this pressure, we obtain a median cloud column mass of only  $\sim 0.98 \times 10^{-2}$, or only $\sim$ 3\% of the total cloud column.

In short, because we are unable to observe the cloud base, we are only permitted to observe a small fraction of the total cloud mass, which will in reality, be completely unknown.

\subsection{A Hazey Warm-Neptune Example} \label{subsec:neptune}
``Warm-Neptunes'' represent a bulk of the transiting exoplanet population \citep{Batalha2014} and are expected to present a diverse range of composition, likely possessing high ($>50\times$ Solar). \textit{HST} transmission spectrum observations \citep{Knutson2014b} suggest that these worlds may possess opaque, high altitude hazes.  These objects are cool enough ($<\ \sim$1000K) that it is not unreasonable to expect the presence of hydrocarbon hazes resulting from methane/ammonia photochemistry (e.g. \citealt{Yung1984, Miller-Ricci Kempton2012, He2018}). However, the detailed mechanisms and chemical-kinetics by which such hazes can form at these atmospheric conditions, or even their composition, as well as the efficiency at which photochemistry can drive their production rate, are largely uncertain (e.g., \citealt{Morley2013}, \citealt{Kawashima2018, Kawashima2019} ). The aim here is to explore possible constraints on the haze properties within a simple, phenomenological parameterization. Our haze model for forward model and retrieval assumes a mixture of ``tholin'' and ``hexene'' (hence, optical properties) and has only four parameters (\S~ \ref{subsubsec:hazemodel}): the haze mixing ratio, the mean particle size at haze base, the haze base pressure, and the tholin fraction (hexene is assumed as the remainder). Figure \ref{fig:specs}c shows the spectrum constructed with this parameterization. It is clearly dominated by the haze contribution.

\begin{figure*}[ht!]
\centering
\includegraphics[scale=0.4]{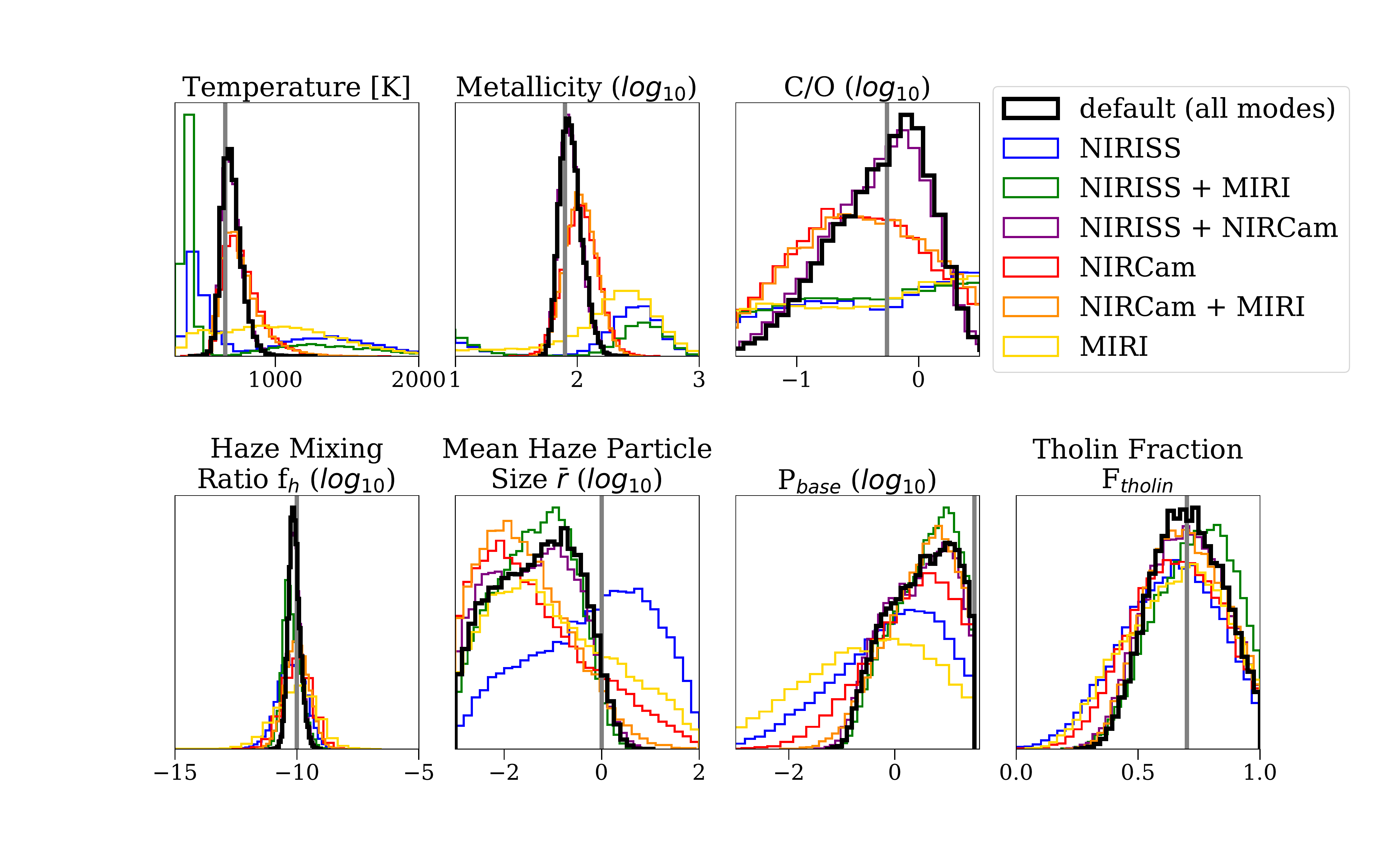}
\caption{Summary posterior probability distributions retrieved in the generic warm-Neptune (GJ 1214b) hazey scenario as a function of instrument mode combination (similar to Figure \ref{fig:modes}). The solid gray lines indicate the truth values (2nd column, Table \ref{tab:values}). The combinationo of NIRISS and NIRCam provide constraints as good as using all three modes.  The 3-5$\mu$m spectral region, covered by NIRCam, is critical to constraining gas phase composition and haze properties due to the simulatenious presence of a strong CO$_2$ feature and haze C-H stretch features. \label{fig:modes_neptune}}
\end{figure*}

Fig. \ref{fig:modes_neptune} summarizes the constraints and their dependence on different instrument mode combinations. When utilizing all modes, the temperature, metallicity, and haze mixing ratio (f$_h$, droplet mixing ratio at base) are tightly constrained -- e.g., bounded.  The C/O ratio, however, is poorly constrained. Although the CH$_4$ features still appear at $\sim$ 3.3 $\mu$m, $\sim$ 7.7 $\mu$m and CO$_2$ feature at $\sim$ 4.2 $\mu$m -- permitting a constraint on metallicity, the H$_2$O features are masked by the featureless haze signature at short wavelengths (Figure \ref{fig:specs}c), prohibiting leverage on the C/O ratio.   f$_{h}$ and F$_{tholin}$ (tholin ``fraction'') are well constrained due to the sensitivity of the 2 - 7 $\mu$m range to the differing C-H stretch features present in the hexene vs. tholin particulates.

Note that the haze base is set deep in the atmosphere (10$^{1.5}$ bar), therefore we are observing only the very upper portion of the haze profile ($<$ 10$^{-4}$ bar).  This means we are sensitive to only the smallest particles within our parameterization, consistent with full microphysical particle growth models \cite{Kawashima2018}.    Though we were are only ``sensing'' a small population of particles within the cloud, because of the integrated haze model parameterization, we are able to constrain f$_{h}$ due to the senstitivity of the entire haze vertical profile to the mixing ratio.  This parameter is constrained over a wide range of instrument modes as ``shifting'' the whole profile (by changing abundance) influences the spectral slope/features at all wavelengths. 

Only an upper limit on the haze base modal particle size can be inferred from most instrument mode combinations.  Larger particle sizes are ruled out as the increase in a larger particle size population aloft would ``flatten'' the spectrum and reduce the presence of the C-H resonance features. It is difficult to rule out smaller mean particle sizes ($< 0.1\ \mu$m) as that population of particles is already present aloft -- well mixed -- and thus decreasing the modal particle size at the base does little to significantly change their vertical distribution via equation \ref{eq3}.  NIRISS alone provides little constraint on the modal particle size. The reducing of posteriors at the edges of the prior range are simply side effects the ellipsoidal sampling of the nested sampling method adopted and therefore should be neglected.

Finally, we find that the haze ``base'' pressure is a ``lower limit'' only. Deep haze bases hardly affect the high-altitude, small particle population, but if it moves towards lower pressures larger sized particles ``move'' into the observable portion of the atmosphere.  The presence of these larger particle sizes, would again, flatten the spectrum and damp the C-H resonance features from 2 - 7 $\mu$m. 

Within this particular planetary scenario, we find that NIRCam is key as it encapsulates the strong, metallicity sensitive, CO$_2$ feature that is able to present itself above the obscuring haze.  Failure to include NIRCam (or equivalently, NIRSpec -- any 3 - 5 $\mu$m coverage) results in bi-modal metallicity and temperature constraints and even more poorly constrained C/O ratios   We also find that the combination of NIRISS and NIRCam provide effectively all of the constraint, without the need for MIRI in this scenario (e.g., the black and purple histograms nearly overlap).   All mode combinations are able to provide an at least 2-orders-of-magnitude constraint on f$_{h}$ and a reasonable grasp on the tholin fraction/composition.  This is largely because the mixing ratio parameter effectively controls the degree of gas-phase feature spectral muting, of which all modes are sensitive.

\section{Discussions} \label{sec:discus}

\subsection{Implications for Future Observations}
The transmission signatures resulting from the condensate cloud parameterizations used here generally include a shallow slope/flattening in the optical and near-IR range, and possibly some weak resonance features in the mid-IR range.  As was mentioned in \ref{subsubsec:noise}, the observed steep scattering slopes reported in the literature at $< 0.8\ \mu$m cannot be explained by ``condensate base'' motivated models (e.g., the A \& M and integrated cloud models), but rather could be due to a number of other phenomenon, including star-spot contamination (\citealt{McCullough2013, Oshagh2014, Rackham2017, Rackham2018}), a strong modal particle size gradient with altitude due to high altitude in-situ formation \citep{Pont2013, Kawashima2019}, or simply an atmosphere scale height increasing with altitude (caused by temperature inversion, low gravity, or decreasing mean molecular weight at high altitudes, etc.)


Nevertheless, in this study we have shown that transmission spectra at longer wavelengths ($> 2.0\ \mu$m) are crucial for the characterization of cloudy/hazy atmospheres. 
The prominent CO feature (good indicator of metallicity) and signatures from C-H, N-H bonds from hydrocarbon haze are mostly located in the wavelength range covered by NIRCam/NIRSpec. Resonance features from mineral condensates (e.g. MgSiO$_3$, Mg$_2$SiO$_4$, Fe$_2$O$_3$, FeSiO$_3$) generally locate in the mid-IR range, covered by MIRI. 


\subsection{Limitations of This Study}
While we have explored the possible constraints and biases on cloud properties within a few specific parameterizations, we freely admit that clouds are undoubtedly complex and we most certainly did not capture the nearly infinite plausible cloud scenarios.  Here we identify some ``known'' weaknesses/caveats with our current investigation.

Firstly, we have only considered a single condensate, MgSiO$_3$ -- anticipated to be a major condensate in the  1100 - 1500 K temperature regime. In reality, numerous other condensates like Mg$_2$SiO$_4$, Fe(L), MnS, Na$_2$S, KCl, Al$_2$O$_3$, etc.  are likely to coexist (as well has spatial in-homogeneities) depending on the exact thermal structure and/or the role of vertical mixing and/or zonal atmospheric dynamics.  Many of these properties will be highly degenerate (e.g., is it even possible to retrieve cloud base pressure and condensate mixing ratio for tens of possible condensate species at once?) and disentangling unique species-specific cloud signatures will likely prove challenging

Ultimately, the practicality of inclusion of more sophisticated cloud models within retrieval frameworks will depend upon their overall computational demand and/or number of free parameters.  It depends upon what the specific goals are. If the goal is to learn about fundamental cloud physics, then certainly efforts in developing more sophsticated cloud parameterizations, motivated by more in depth self-consitent 3D studies are invaluable.  However, if the goal is to ``marginalize out the clouds'' so as to not bias other desireable atmospheric quantities (e.g., composition, temperature), we've shown here, that simply including this reasonable set of parameterizations (the A \& M cloud model) appears to be good enough.

\section{Summary}   \label{sec:sum}
Within an atmospheric retrieval framewrok, we have explored the degree to which atmospheric and cloud/haze properties can be constrained from transmission spectra as observed under various \textit{JWST} scenarios under several different cloud parameterizations using a generic hot-Jupiter and a warm-Neptune as test cases. 



From this investigation, we have come to the following major conclusions:

\begin{enumerate}
\item Under a realistic condensate cloud model assumption, transmission spectra are likely to feature a shallow slope truncating the gas features across the visible and near-IR wavelength range and small resonance features at longer wavelengths (similar to \cite{Molliere2017}). 

\item The \textit{JWST} transmission spectra of cloudy atmospheres have the potential to constrain the key atmosphere parameters (temperature, metallicity and C/O ratio), the key cloud parameters from the fiducial Ackerman \& Marley cloud model ($f_{sed}$, $K_{zz}$ and the cloud base pressure). Without viewing a cloud base in the spectrum, we are only able to constrain a lower limit of the cloud abundance and the observable upper part of the cloud mass, even with the most precise data we have probed. 

\item Higher spectral precision generally results in tighter constraints on the atmospheric temperature, metallicity, and C/O ratio.  In some cases, however, increased spectral precision, (covering at least a factor of $\sim$4 in precision) does not improve constraints on f$_{cond}$ or P$_{base}$.  Increasing wavelength coverage is a more fruitful way to boost both cloud and composition constraints. In the generic hot-Jupiter scenario, the 4-5$\mu$m spectral range is critical to constraining the metallicity due to the strong presence of CO \& CO$_2$ features. The NIRISS + NIRCam (0.6 - 5 $\mu$m coverage) combination provides temperature, metallicity, and C/O constraints almost as precise as the full 1-12$\mu$m wavelength range. However, MIRI is critical for constraining cloud properties due to the resonant features at longer wavelengths. Individual instrument modes by themselves are generally less than optimal.

\item A cloud-free atmosphere retrieval results in in biases and poor constraints on the temperature, metallicity, and C/O ratio when applied to cloudy atmosphere. All four cloud models investigated in this study (the power law haze + gray cloud model, the one-size cloud model, the integrated cloud model, and the Ackerman and Marley cloud model) do not introduce serious biases in the retrieved temperature, metallicity, or C/O -- just so long as a cloud model is used.

\item Degeneracies exist between the cloud parameters in the Ackerman \& Marley model, including between the $f_{sed}$ and $K_{zz}$, and between f$_{cond}$ and P$_{base}$.  The former stems from their intrinsic physical correlation in controlling particle size. The latter comes from our lack of knowledge of the total amount of cloud due to our lack of knowledge of the cloud base. Higher spectral precision does little to alleviate this degeneracy.

\item The transmission spectrum of a high metallicity warm-Neptune with a ``photochemically derived'' haze is able to constrain the atmospheric temperature, metallicity, haze base mixing ratio, and haze composition if the 4-5$\mu$m range is included due to the presense of C-H/N-H stretch features of the haze consitutents and the strong presence of CO$_2$.


\end{enumerate}

\acknowledgments

Acknowledgments --  
We thank the anonymous reviewer for the careful reading and comments which greatly help improve the manuscript. 
We are grateful to Adam Schneider, Ehsan Gharib-Nezhad for helpful comments to improve the manuscript. We thank Diana Powell, Peter Gao, Paul Molli\`ere, Laura Kreidberg, Mark Marley, Jonathan Fortney, Ryan MacDonald, Lorenzo Pino for meaningful discussions.  We also thank Hannah Wakeford for providing us with ascii tables of the indicies of refraction presented in \cite{Wakeford2015}. The authors acknowledge Research Computing at Arizona State University for providing High Performance Computing (HPC) resources that have contributed to the research results reported within this paper. This work benefited from the 2018 Exoplanet Summer Program in the Other Worlds Laboratory (OWL) at the University of California, Santa Cruz, a program funded by the Heising-Simons Foundation. This work was supported by the NASA Exoplanet Research Program award NNX17AB56G.

%


\facility{\textit{JWST, HPC Agave cluster \& Saguaro cluster at Arizona State University}}

\software{PyMultiNest \citep{Buchner2014}, PyMieCoated (https://github.com/jleinonen/pymiecoated/),
Chimera (https://github.com/ExoCTK/chimera)}


\newpage
\begin{thebibliography}{}

\bibitem[Ackerman \& Marley(2001)]{Ackerman2001} Ackerman, A. S., \& Marley, M. S.\ 2001, \apj, 556,872
\bibitem[Adams et al.(2019)]{Adams2019} Adams, D., Gao, P., de Pater, I., \& Morley, C. V. 2019.\ \apj, 874, 61
\bibitem[Allard et al.(2012)]{Allard2012} Allard, F., Homeier, D., Freytag, B., \& Sharp, C. M.\ 2012, EAS Publications Series, Vol. 57, EAS Publications Series, ed. C. Reyl\'{e}, C. Charbonnel, \& M. Schultheis, 3-43
\bibitem[Batalha(2014)]{Batalha2014} Batalha, N. M.\ 2014, Proceedings of the National Academy of Sciences, 111, 35, 12647-12654
\bibitem[Benneke \& Seager(2013)]{Benneke2013} Benneke, B., \& Seager, S.\ 2013, \apj, 778, 153
\bibitem[Blecic et al.(2017)]{Blecic2017} Blecic, J., Dobbs-Dixon, I., Greene, T.\ 2017, \apj, 848, 127
\bibitem[Brown(2001)]{Brown2001} Brown, T. M.\ 2001, \apj, 553, 1006
\bibitem[Buchner et al.(2014)]{Buchner2014} Buchner, J., Georgakakis, A., Nandra, K., et al.\ 2014, \aa, 564, A125 
\bibitem[Burrows et al.(2006)]{Burrows2006} Burrows, A., Sudarsky, D., \& Hubeny, I.\ 2006, \apj, 640, 1063
\bibitem[Caldas et al.(2019)]{Caldas2019} Caldas, A., Leconte, J., Selsis, F., et al.\ 2019, \aa, 623, A161
\bibitem[Charnay et al.(2018)]{Charnay2018} Charnay, B., Bézard, B., Baudino, J. -L., Bonnefoy, M., Boccaletti, A., Galicher, R.\ 2018, \apj, 854, 172
\bibitem[Crossfield et al.(2013)]{Crossfield2013} Crosfield, I. J. M., Barman, T., Hansen, B. M. S., \& Howard, A. W.\ 2013, \aa, 559, A33
\bibitem[Feng et al.(2016)]{Feng2016} Feng, Y. K., Line, M. R., Fortney, J. J., et al.\ 2016, \apj, 829, 52
\bibitem[Feng et al.(2018)]{Feng2018} Feng, Y. K., Robinson, T. D., Fortney, J. J., Lupu, R. E., Marley, M. S., Lewis, M. K., Macintosh, B., Line, M. R.\ 2018, \aj, 155, 200
\bibitem[Feroz \& Hobson(2008)]{Feroz2008} Feroz, F., \& Hobson, M. P.\ 2008, \mnras, 384, 449
\bibitem[Fortney(2005)]{Fortney2005} Fortney, J. J.\ 2005, \mnras, 364, 649
\bibitem[Fraine et al.(2014)]{Fraine2014} Fraine, J., Deming, D., Benneke, B., Knutson, H., Jordán, A., Espinoza, N., Madhusudhan, N., Wilkins, A., Todorov, K.\ 2014, \nat, 513, 526
\bibitem[Freedman et al.(2008)]{Freedman2008} Freedman, R. S., Marley, M. S., \& Lodders, K.\ 2008, \apjs, 174, 504
\bibitem[Freedman et al.(2014)]{Freedman2014} Freedman, R. S., Lustig-Yaeger, J., Fortney, J. J., Lupu, R. E., Marley, M. S., Lodders, K.\ 2014, \apjs, 214, 25 
\bibitem[Gao et al.(2018)]{Gao2018a} Gao, P., Marley, M. S., \& Ackerman, A. S.\ 2018, \apj, 885, 86
\bibitem[Gao \& Benneke(2018)]{Gao2018b} Gao, P., \& Benneke, B. 2018.\ \apj, 863, 165
\bibitem[Gorden \& McBride(1996)]{Gorden1996} Gordon, S., McBride, B. J., \& NASA Tech. Info. Program.\ 1996, Computer program for calculation of complex chemical equilibrium compositions and applications, National Aeronautics and Space Administration, Office of Management, Scientific and Technical Information Program
\bibitem[Greene et al.(2016a)]{Greene2016a} Greene, T. P.,	Chu, L., Egami, E., et al.\ 2016a, Proc. SPIE, 9904, 99040E
\bibitem[Greene et al.(2016b)]{Greene2016} Greene, T. P., Line., M. R., Montero, C., Fortney, J. J., Lustig-Yaeger, J., \& Luther K.\ 2016b, \apj, 817, 17
\bibitem[Guillot(2010)]{Guillot2010} Guillot, T.\ 2010, \aa, 520, A27
\bibitem[He et al.(2018)]{He2018} He, Chao, Hörst, Sarah M., Lewis, Nikole K., et al.\ 2018, \aj, 156, 38
\bibitem[Helling et al.(2008a)]{Helling2008a} Helling, C., Dehn, M., Woitke, P., \& Hauschildt, P. H.\ 2008a, \apjl, 675, L105
\bibitem[Helling et al.(2008b)]{Helling2008b} Helling, C., Woitke, P., \& Thi, W.-F.\ 2008b, \aa, 485, 547
\bibitem[Helling et al.(2016)]{Helling2016} Helling, C., , G., Dobbs-Dixon, I., et al.\ 2016, \mnras, 460, 855
\bibitem[Huitson et al.(2012)]{Huitson2012} Huitson, C. M., Sing, D. K., Vidal-Madjar, A. et al.\ 2012, \mnras, 422, 2477
\bibitem[Iyer et al.(2016)]{Iyer2016} Iyer, A. R., Swain, M. R., Zellem, R. T., et al.\ 2016, \apj, 823, 109
\bibitem[Kawashima \& Ikoma(2018)]{Kawashima2018} Kawashima, Y., \& Ikoma, M.\ 2018, \apj, 853, 7
\bibitem[Kawashima \& Ikoma(2019)]{Kawashima2019} Kawashima, Y., \& Ikoma, M.\ 2019, arXiv:1904.09986 
\bibitem[Knutson et al.(2014a)]{Knutson2014a} Knutson H. A., Benneke B., Deming D., Homeier D.\ 2014a, \nat, 505, 66
\bibitem[Knutson et al.(2014b)]{Knutson2014b} Knutson H. A. et al.\ 2014b, \apj, 794, 155
\bibitem[Kreidberg et al.(2014)]{Kreidberg2014} Kreidberg, L., Bean, J. L., D'esert, J.-M., et al.\ 2014, \nat, 505, 69
\bibitem[Kreidberg et al.(2015)]{Kreidberg2015} Kreidberg, L., Line, M. R., Bean, J. L., et al.\ 2015, \apj, 814, 66
\bibitem[Kreidberg et al.(2018)]{Kreidberg2018} Kreidberg, L., Line, M. R., Thorngren, D., Morley, C. V., \& Stevenson, K. B.\ 2018, \apjl, 858, L6
\bibitem[Lacis \& Oinas(1991)]{Lacis1991} Lacis, A. A., \& Oinas, V.\ 1991, \jgr, 96. 9027
\bibitem[Lavvas \& Koskinen(2017)]{Lavvas2017} Lavvas, P., \& Koskinen, T.\ 2017, \apj, 847, 32
\bibitem[Lecavelier des Etangs et al.(2008)]{desEtangs2008} Lecavelier des Etangs, A., Pont, F., Vidal-Madjar, A., Sing, D.\ 2008, \aa, 481
\bibitem[Lee et al.(2014)]{Lee2014} Lee J.-M., Irwin P. G. J., Fletcher L. N., Heng K., Barstow J. K.\ 2014, \apj, 789, 14
\bibitem[Lee et al.(2015)]{Lee2015} Lee, G., Helling, C., Dobbs-Dixon, I., \& Juncher, D.\ 2015, \aa, 580, A12
\bibitem[Lee et al.(2016)]{Lee2016} Lee, G., Dobbs-Dixon, I., Helling, C., Bognar, K., \& Woitke, P.\ 2016, \aa, 594, A48
\bibitem[Line et al.(2013)]{Line2013} Line, M. R., Wolf, A., Zhang, X., et al.\ 2013b, \apj, 775, 137
\bibitem[Line \& Parmentier(2016)]{Line2016} Line, M. R., \& Parmentier, V.\ 2016, \apj, 820, 78
\bibitem[Lines et al.(2018)]{Lines2018} Lines, S., Manners, J., Mayne, N. J. et al.\ 2018, \mnras, 481, 194
\bibitem[Lupu et al.(2016)]{Lupu2016} Lupu, R. E., Marley, M. S., Lewis, N.\ 2016, \aj, 152, 217
\bibitem[Madhusudhan et al.(2011)]{Madhusudhan2011} Madhusudhan, N., Burrows, A., \& Currie, T.\ 2011, \apj, 737, 34
\bibitem[Marley et al.(2010)]{Marley2010} Marley, M. S., Saumon, D., \& Goldblatt, C.\ 2010, \apj, 723, L117
\bibitem[McCullough et al.(2013)]{McCullough2013} McCullough, P. R., Crouzet, N., Deming, D., Madhusudhan, N.\ 2014, \apj, 791, 55
\bibitem[Miller-Ricci Kempton et al.(2012)]{Miller-Ricci Kempton2012} Miller-Ricci Kempton, E., Zahnle, K., \& Fortney, J. J.\ 2012, \apj, 745, 3
\bibitem[Molli\`{e}re et al.(2017)]{Molliere2017} Molli\`{e}re, P., van Boekel, R., Bouwman, J., et al.\ 2017, \aap, 605, C3
\bibitem[Morley et al.(2012)]{Morley2012} Morley, C. V., Fortney, J. J., Marley, M. S., et al.\ 2012, \apj, 756, 172
\bibitem[Morley et al.(2013)]{Morley2013} Morley, C. V., Fortney, J. J., Kempton, E. M. -R., Marley, M. S., Visscher, C., Zahnle, K.\ 2013, \apj, 775, 33
\bibitem[Morley et al.(2014)]{Morley2014} Morley, C. V., Marley, M. S., Fortney, J. J., et al.\ 2014, \apj, 787, 78
\bibitem[Morley et al.(2015)]{Morley2015} Morley, C. V., Fortney, J. J., Marley, M. S., et al.\ 2015, \apj, 815, 22
\bibitem[Moses et al.(2011)]{Moses2011} Moses, J. I., Visscher, C., Fortney, J. J., et al.\ 2011, \apj, 737, 15
\bibitem[Ohno \& Okuzumi(2018)]{Ohno2018} Ohno, K., \& Okuzumi, S.\ 2018, \apj, 859, 34
\bibitem[Ormel \& Min(2019)]{Ormel2019} Ormel, C. W., \& Min, M. 2019.\ \aap, 622, A121
\bibitem[Oshagh et al.(2014)]{Oshagh2014} Oshagh, M., Santos, N. C., Ehrenreich, D., et al.\ 2014, \aap, 568, A99
\bibitem[Parmentier et al.(2013)]{Parmentier2013} Parmentier, V., Showman, A. P., \& Lian, Y.\ 2013, \aap, 558 A91
\bibitem[Pinhas \& Madhusudhan(2017)]{Pinhas2017} Pinhas, A., \& Madhusudhan, N.\ 2017, \mnras, 471, 4355
\bibitem[Pinhas et al.(2018)]{Pinhas2018} Pinhas, A., Rackham, B. V., Madhusudhan, N., \& Apai, D.\ 2018, \mnras, 480, 5314
\bibitem[Pont et al.(2008)]{Pont2008} Pont F., Knutson H., Gilliland R. L., Moutou C., Charbonneau D.\ 2008, \mnras, 385, 109
\bibitem[Pont et al.(2013)]{Pont2013} Pont F., Sing D. K., Gibson N. P., Aigrain S., Henry G., Husnoo N.\ 2013, \mnras, 432, 2917
\bibitem[Powell et al.(2018)]{Powell2018} Powell, D., Zhang, X., Gao, P., Parmentier, V.\ 2018, \apj, 860, 18
\bibitem[Rackham et al.(2017)]{Rackham2017} Rackham B., et al.\ 2017, \apj, 834, 151
\bibitem[Rackham et al.(2018)]{Rackham2018} Rackham B. V., Apai D., Giampapa M. S.\ 2018, \apj, 853, 122
\bibitem[Rocchetto et al.(2016)]{Rocchetto2016} Rocchetto, M., Waldmann, I. P., Venot, O., Lagage, P. -O., \& Tinetti, G.\ 2016, \apj, 833, 120
\bibitem[Schlawin et al.(2017)]{Schlawin2017} Schlawin, E., Rieke, M., Leisenring, J., et al. 2017, PASP, 129, 015001
\bibitem[Schlawin et al.(2018)]{Schlawin2018} Schlawin, E., Greene, T. P., Line, M. R., Fortney, J. J., \& Rieke, M.\ 2018, \aj, 156, 40
\bibitem[Seager \& Sasselov(2000)]{Seager2000} Seager, S., \& Sasselov, D. D.\ 2000, \apj, 537, 916
\bibitem[Sing et al.(2016)]{Sing2016} Sing D. K. et al.\ 2016, \nat, 529, 59
\bibitem[Skilling(2004)]{Skilling2004} Skilling, J.\ 2004, AIP Conference Proceedings, 735, 395
\bibitem[Swain et al.(2008)]{Swain2008} Swain, M. R., Vasisht, G,, \& Tinetti, G.\ 2008, \nat, 452, 329
\bibitem[Tinetti et al.(2012)]{Tinetti2012} Tinetti, G., Beaulieu, J. P., Henning, T., et al.\ 2012, Experimental Astronomy, 34, 311
\bibitem[Tsiaras et al.(2018)]{Tsiaras2018} Tsiaras, A., Waldmann, I. P., Zingales, T., et al.\ 2018, \aj, 155, 4
\bibitem[Tsuji(2002)]{Tsuji2002} Tsuji, T. 2002, ApJ, 575, 264
\bibitem[Vahidinia et al.(2014)]{Vahidinia2014} Vahidinia, S., Cuzzi, J. N., Marley, M., \& Fortney, J.\ 2014, \apj, 789, L11
\bibitem[Wakeford \& Sing(2015)]{Wakeford2015} Wakeford H. R., Sing D. K.\ 2015, \aa, 573, A122
\bibitem[Wakeford et al.(2017)]{Wakeford2017} Wakeford H. R., Visscher C., Lewis N. K., Kataria T., Marley M. S., Fortney J. J., Mandell A. M.\ 2017, \mnras, 464, 4247
\bibitem[Yung et al.(1984)]{Yung1984} Yung, Y. L., Allen, M., \& Pinto, J. P.\ 1984, \apjs, 55, 465




\end{thebibliography}
\end{document}